# Electromechanical Detection in Scanning Probe Microscopy: Tip Models and Materials Contrast

Eugene A. Eliseev[*]

Institute for Problems of Materials Science, National Academy of Science of Ukraine,

3, Krjijanovskogo, 03142 Kiev, Ukraine

Sergei V. Kalinin[†] and Stephen Jesse

Materials Science and Technology Division, Oak Ridge National Laboratory

Oak Ridge, TN 37831

Svetlana L. Bravina

Institute of Physics, National Academy of Science of Ukraine,

46, pr. Nauki, 03028 Kiev, Ukraine

Anna N. Morozovska

V.Lashkaryov Institute of Semiconductor Physics, National Academy of Science of Ukraine,

41, pr. Nauki, 03028 Kiev, Ukraine

---

[*] Corresponding author, eliseev@i.com.ua

[†] Corresponding author, sergei2@ornl.gov

**Abstract**

The rapid development of nanoscience and nanotechnology in the last two decades was stimulated by the emergence of scanning probe microscopy (SPM) techniques capable of accessing local material properties, including transport, mechanical, and electromechanical behavior on the nanoscale. Here, we analyze the general principles of electromechanical probing by piezoresponse force microscopy (PFM), a scanning probe technique applicable to a broad range of piezoelectric and ferroelectric materials. The physics of image formation in PFM is compared to Scanning Tunneling Microscopy and Atomic Force Microscopy in terms of the tensorial nature of excitation and the detection signals and signal dependence on the tip-surface contact area. It is shown that its insensitivity to contact area, capability for vector detection, and strong orientational dependence render this technique a distinct class of SPM. The relationship between vertical and lateral PFM signals and material properties are derived analytically for two cases: transversally-isotropic piezoelectric materials in the limit of weak elastic anisotropy, and anisotropic piezoelectric materials in the limit of weak elastic and dielectric anisotropies. The integral representations for PFM response for fully anisotropic material are also obtained. The image formation mechanism for conventional (e.g., sphere and cone) and multipole tips corresponding to emerging shielded and strip-line type probes are analyzed. Resolution limits in PFM and possible applications for orientation imaging on the nanoscale and molecular resolution imaging are discussed.

# I. Introduction

Rapid progress in nanoscience and nanotechnology in the last two decades has been stimulated by and also necessitates further development of tools capable of addressing material properties on the nanoscale.[1] Following the development of Scanning Tunneling Microscopy[2] (STM) and Atomic Force Microscopy[3] (AFM) techniques that allowed visualizing and manipulating matter on the atomic level, a number of force- and current based Scanning Probe Microscopy (SPM) techniques were developed to address properties such as conductance, elasticity, adhesion, etc. on the nanoscale.[4,5] Currently, the central paradigms of existing SPM techniques are based on detection of current induced by the bias applied to the probe, and cantilever displacement induced by a force acting on the tip. The third possibility, electromechanical detection of surface displacements due to piezoelectric and electrostrictive effects induced by bias applied to the probe tip, is realized in Piezoresponse Force Microscopy (PFM). These three detection mechanisms can be implemented on both SPM and nanoindentor based-platforms. Finally, detection of current induced by force applied to the probe is limited by the smallness of the corresponding capacitance and has not been realized in SPM. However, such measurements have been realized on nanoindentor-based platforms.[6,7]

PFM was originally developed for imaging, spectroscopy, and modification of ferroelectric materials with strong electromechanical coupling coefficients (20–2000 pm/V).[8,9,10] The ability to measure vertical and lateral components of the electromechanical response vector, perform local polarization switching, and measure local hysteresis loops (PFM spectroscopy) has attracted broad attention to this technique and resulted in a rapidly increasing number of publications.[11,12] Currently, PFM is one of the most powerful tools for nanoscale characterization of ferroelectric materials. However, until recently, it was believed that PFM was limited to ferroelectric materials, representing a relatively minor class of inorganic materials and, with few exceptions (e.g., polyvinilidendifluoride and its copolymers) non-existent in macromolecular materials and biopolymers.

However, piezoelectric coupling between electrical and mechanical phenomena is extremely common in inorganic materials (20 out of 32 symmetry classes are piezoelectric) and ubiquitous in biological polymers due to the presence of polar bonds and optical activity. Hence, PFM is a natural technique for high-resolution imaging of these systems. One of the limitations in PFM of such materials is low electromechanical coupling coefficients, typically 1–2 orders of magnitude below that of ferroelectrics. However, the high vertical resolution, inherent to all SPM techniques, combined with large (~1–10 Vpp) modulation amplitudes allows local measurement of electromechanical coupling even in materials with small piezoelectric coefficients, e.g., III–V nitrides[13] and biopolymers.[14,15] In fact, the primary limitations in PFM imaging of weakly piezoelectric materials are not the smallness of the corresponding response magnitude, but the linear contribution to the PFM contrast due to capacitive tip-surface forces and the inability to use standard phase-locked loop based circuitry for resonance enhancement of weak electromechanical signal.[16] However, both of these limitations can be circumvented by electrically shielded probes, imaging in liquid environment,[17] and improved control and signal acquisition routines.

Finally, we note that while piezoelectricity, similar to elasticity and the dielectric constant, is a bulk property defined only for an atomically large (many unit cells) volume of material, the electromechanical coupling *per se* exists down to a single polar bond level.[18] Hence theoretically, electromechanical properties can be probed on molecular and atomic levels. As a consequence of piezoelectricity in polar bonds, all polar materials possess piezoelectric properties, unless forbidden by lattice symmetry. Symmetry breaking at surfaces and interfaces should give rise to piezoelectric coupling even in non-polar materials, and a number of novel electromechanical phenomena, including surface piezoelectricity and flexoelectricity, have been predicted.[19]

To summarize, electromechanical coupling is extremely common on the nanoscale, it is manifest on the single molecule level, and novel forms are enabled at surfaces and in nanoscale systems unconstrained by bulk symmetry. PFM is a natural tool to address these phenomena, in turn necessitating the understanding of the relationship between signal formation mechanisms, material properties, and tip parameters. The relationship between the surface and tip displacement amplitudes,[20] cantilever dynamics in PFM,[21] and mechanisms for electrostatic force contribution[22,23] have been analyzed in detail. However, due to its inherent complexity, voltage-dependent tip-surface contact mechanics, the key element that relates applied modulation and measured response, is available only for a limited class of transversally-isotropic piezoelectric materials and simple tip geometries.

Here, we analyze the basic physics of PFM in terms of the tensorial nature of the response, the signal dependence on contact area, materials properties contributing to the signal, and signal-distance dependence. The image formation mechanisms in current- and force-based scanning probe microscopies is analyzed in terms of the tensor nature of the excitation and detection signal's sensitivity to contact area in Section II. The relationship between the PFM signal and material properties and tip geometry is analyzed using a linearized decoupled Green's function approach in Section III. Orientation imaging by PFM and implications for high-resolution electromechanical imaging are discussed in Section IV.

## II. Classification of SPMs

In this section, current- and force-based SPM techniques are discussed in terms of the tensor nature of the measured signal and the dependence of the measured signal on the contact area. This means of classification is complementary to the widely used classification of the techniques based on the apparatus, in which microscopes using probes with the cantilever- or tuning-fork based displacement detection system are referred to as AFMs and systems employing current detection through etched metal probe are referred to as STMs. By addressing the physical mechanisms behind the contrast as opposed to instrumental platform, mechanism based classification clarifies the role of individual interactions and elucidates strategies for further technique development, such as applicability of resonant enhancement, sensitivity to topographic cross-talk, etc.

In a general local probe experiment implemented on either a SPM or nanoindentor platform, there are two independently controlled external variables, namely probe bias and indentation force (Fig. 1). Two independently detected parameters are cantilever deflection (or changes in the dynamic mechanical characteristics of the system) and probe current. In STM and conductive AFM, current induced by the bias applied to the tip is detected and is used as a feedback signal for tracing topography (STM) or local conductivity measurements (cAFM). In AFM and related techniques, including non-contact and intermittent contact AFM, atomic force acoustic microscopy, etc. the cantilever displacement induced by a force applied to the probe is detected. The force can also be of an electrostatic and magnetic nature, providing the basis of electrostatic and magnetic force microscopies. The response in this case scales reciprocally with cantilever spring constant. In PFM, mechanical displacement induced by an electric bias applied to the tip is detected and response is independent on cantilever spring constant. The reverse mechanism, detection of the current induced by the force, is limited by the smallness of the corresponding capacitance in an SPM experiment, and can be implemented on nanoindentator platforms where the contact areas are significantly larger (contact mode),[6,7] or by using especially large radius of curvature tips (non-contact mode). From these considerations, the signal formation mechanism in PFM is distinctly different from that in either AFM or STM and conductive AFM.

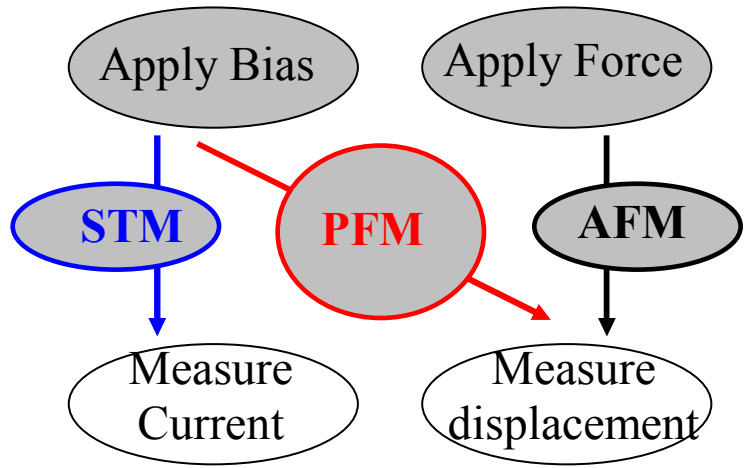

**Fig. 1.** Classification of SPM techniques based on the detection mechanism.

To establish the capacity of an SPM technique for quantitative measurements, we consider the signal formation mechanism in terms of the tensor nature of the input and output signals and the dependence of the signal on contact area. Here, we assume that the measurements are performed in the contact regime, in which the mechanical contact between the tip and the surface is well-defined. The necessary conditions for quantitative measurements by SPM are either (a) the contact area and tip geometry and properties are known or are easy to calibrate or (b) signal is insensitive to contact area and tip properties. The first approach is adopted in nanoindentation, where calibration of the indentor shape is the crucial step of a quantitative experiment.[24] With few exceptions, tip shape calibration has not yet become routine in AFM-based measurements.[25,26] In addition, tip state often changes in STM and AFM,[27] requiring the development of rapid characterization methods. Therefore, of interest are SPM techniques in which this image formation mechanism is such that the signal does not depend on the contact area, either due to fundamental physics of tip-surface interactions or because the contact area is confined to single atom or molecule. The dependence of the signal on tip-surface separation is a key factor in determining the spatial resolution in the technique. Finally, the tensorial nature of the signal determines the number of independent data channels that can be accessed in the ideal experiment.

In current-based techniques such as STM, the input signal, i.e., bias applied to the probe, is a scalar quantity. Electrical current is a vector having three independent components, but, because of the point contact geometry inherent to SPM, the detected signal is current magnitude, i.e., a scalar quantity. Hence, STM and conductive AFM signal relates two scalar quantities and thus is a scalar. The relationship between the excitation and measured signals is

$$I = \Sigma V, \tag{1}$$

where $\Sigma$ is local conductance determined by material properties, contact area, tip-surface separation, and tip geometry. Note that here and everywhere we consider the case of semi-infinite uniform 3D material, for which the response to the point excitation is determined by local properties only and is independent on the boundary conditions.

The local conductance depends on contact area. In the ohmic regime, the spreading conductance scales as $\Sigma \sim a^1$.[28] For small contact areas comparable to the Fermi wavelength of the electrons, the crossover to the Sharwin regime occurs, $\Sigma \sim a^2$.[29] Finally, in the case when the contact is confined to single atom in a tunneling regime, $\Sigma \sim a^0$. The classical regime corresponds to operation in Scanning Spreading Resistance microscopy and conductive AFM. Here, conductivity strongly depends on the local carrier concentration that typically varies over many orders of magnitude. Hence, relatively small changes in contact geometry due to surface topography at strong indentation forces are negligible. However, c-AFM of conductive topographically inhomogeneous materials is likely to provide increased signal at pores, grain boundaries, etc., due to changes in effective contact area. At the same time in atomic-resolution STM, the contact is confined to a single atom and is hence much better defined, suggesting a greater potential of this technique for quantitative measurements.

In force-based techniques, such as AFM, both the excitation signal (i.e., force) and the response signal (i.e., displacement) are vectors. Hence, the AFM signal is a rank two tensor. In the coordinate system aligned with cantilever, where the 1-axis is oriented in the surface plane along the cantilever, the 2-axis is in-plane and perpendicular to the cantilever, and 3-axis is the surface normal, the signal formation mechanism can be represented as

$$\begin{pmatrix} w_1 \\ w_2 \\ w_3 \end{pmatrix} = \begin{pmatrix} a_{11} & a_{12} & a_{13} \\ a_{21} & a_{22} & a_{23} \\ a_{31} & a_{32} & a_{33} \end{pmatrix} \begin{pmatrix} F_1 \\ F_2 \\ F_3 \end{pmatrix} \qquad (2)$$

The component of the stiffness tensor $a_{33}$ is probed by conventional vertical AFM and AFAM, while $a_{22}$ is probed in lateral force microscopy. The off-diagonal component, $a_{23}$ and $a_{32,}$ represent coupling between the vertical and lateral signals. However, a conventional cantilever sensor does not allow separation between the normal and longitudinal force components, both of which result in flexural deformation of the cantilever. The lateral component results in cantilever torsion and can be measured independently. Hence, in terms of measured components, the flexural, $fl$, and torsional, $tr$, responses are

$$\begin{pmatrix} fl \\ tr \end{pmatrix} = \begin{pmatrix} \alpha a_{11} + \beta a_{31} & \alpha a_{12} + \beta a_{32} & \alpha a_{13} + \beta a_{33} \\ \chi a_{21} & \chi a_{22} & \chi a_{23} \end{pmatrix} \begin{pmatrix} F_1 \\ F_2 \\ F_3 \end{pmatrix}, \quad (3)$$

where $\alpha$, $\beta$, and $\chi$ are proportionality coefficients dependent on cantilever geometry and calibration. Hence, information on materials response is partially lost in cantilever based experiment. This coupling between longitudinal and normal displacements is a well-recognized problem in AFM, hindering quantitative indentation measurements with standard cantilever sensors.[30,31] A number of attempts to develop 3D force sensors avoiding this limitation have been reported.[32,33]

In the continuum mechanics limit, the contact stiffnesses are proportional to contact radius, $a_{ij} \sim a^1$. This behavior holds down to length scales of a few atoms. When the contact area is single molecule, as in protein unfolding spectroscopy, the effective contact area is constant and $a_{ij} \sim a^0$. Hence, quantitative force measurements are generally limited to the cases when the contact geometry is well characterized, as in nanoindentation techniques, or is weakly dependent on the probe, as in atomic-resolution imaging or molecular unfolding.

PFM employs electromechanical detection. The excitation signal is bias, whereas the electromechanical response of the surface is a vector. Hence, the PFM response is a vector,

$$\begin{pmatrix} w_1 \\ w_2 \\ w_3 \end{pmatrix} = \begin{pmatrix} d_1 \\ d_2 \\ d_3 \end{pmatrix} V. \quad (4)$$

Here, the response components $d_i$ describe the electromechanical coupling in the material in the point contact geometry. In the uniform field case, $(d_1, d_2, d_3) = (d_{34}, d_{35}, d_{33})$, where $d_{ij}$ are longitudinal and shear elements of the piezoelectric constant tensor of the material in the laboratory coordinate system, as discussed in detail elsewhere.[20]

Similarly to AFM, the torsional and flexural components of the cantilever oscillations, rather than the surface displacement components, are measured in PFM experiments, resulting in mixing between longitudinal, $d_1$, and normal, $d_3$, components of the signal. However, due to differences in the signal transduction mechanism between normal and shear surface vibrations and tip motion, the vertical signal can be measured using high frequency excitation. At the same time, both in-plane components of the response vector can be detected by imaging before and after a 90° in-plane rotation of the sample. This approach, while tedious and limited to samples with clear topographic markings necessary for locating same region after rotation, allows the full response vector to be obtained.[34,35] Further progress can be achieved with 3D force probes.[32,33]

The distinctive feature of electromechanical response, distinguishing it from the detection mechanisms in-force and current-based techniques, is the nature of the signal dependence on contact area. From simple dimensionality considerations, the electromechanical response does not depend on contact area, $d_i \sim a^0$. This conjecture in the continuum mechanics limit for transversally isotropic materials is confirmed by rigorous theory.[36,37] On a smaller length scales, a number of extrinsic (e.g., potential drop in the tip-surface gap due to Thomas-Fermi screening in tip metal and non-ferroelectric surface layer) and intrinsic (e.g., when contact area and excitation volume become comparable to the correlation length) mechanisms leading to different scaling of responses with contact area can become important. To the best of the authors' knowledge, the electromechanical response in ferroelectrics as a function of contact area has not, to date, been studied systematically.

However, given that the piezoelectricity in molecular and polar materials originates on a single bond level, for small contact areas the behavior $d_i \sim a^0$ is expected.

To summarize, the image formation mechanism in PFM is distinctly different from conventional force- and current-based SPM techniques. In the classical limit, the signal is independent of the contact area, thus providing a basis for quantitative measurements. All three components of the response can be measured, providing comprehensive information on material properties. The factors combined with the ubiquitous presence of piezoelectricity in inorganic materials, polar polymers, and biopolymers, necessitate a detailed analysis of the image formation mechanism in PFM. Below, we analyze the relationships between material properties and contrast in PFM, i.e., the nature of the responses in Eq. (3) for different tip geometries, and briefly discuss the dependence of the PFM signal on the tip-surface separation distance.

### III. Materials Contrast in PFM

In general, a calculation of the electromechanical response induced by a biased tip requires the solution of a coupled electromechanical indentation problem, currently available only for uniform transversally isotropic case.[36,37] This solution is further limited to the strong indentation case, in which the fields generated outside the contact area are neglected. While this approximation is valid for large contact areas, for small contacts the electrostatic field produced by the part of the tip not in contact with the surface can provide a significant contribution to the electromechanical response. This behavior is similar to the transition from Hertzian contact mechanics valid for macroscopic contacts to Dugdale-Maugis mechanics for nanoscale contacts.[38,39]

An alternative approach for the calculation of the electromechanical response is based on the decoupling approximation. In this case, the electric field in the material is calculated using a rigid electrostatic model (no piezoelectric coupling), the strain or stress field is calculated using constitutive relations for piezoelectric material, and the displacement field is evaluated using the appropriate Green's function for an isotropic or anisotropic solid. This approach is rigorous for the materials with small piezoelectric coefficients. A simple estimation of the decoupling approximation applicability is based on the value of the square of the dimensionless electromechanical coupling coefficients $k_{ij}^2 = (d_{ij})^2 / (s_{jj}\varepsilon_{ii})$ (see Appendix A). For instance for BaTiO$_3$: $k_{15}^2 \approx 0.32$, $k_{31}^2 \approx 0.10$, and $k_{33}^2 \approx 0.31$, for the ceramics PZT6B: $k_{15}^2 \approx 0.14$, $k_{31}^2 \approx 0.02$, and $k_{33}^2 \approx 0.13$, and for a quartz single crystal: $k_{11}^2 \approx 0.01$.

A simplified 1D version of the decoupled model was suggested by Ganpule[40] to account for the effect of 90° domain walls on PFM imaging. A similar 1D approach was adapted by Agronin et al.[41] to yield closed-form solutions for the PFM signal. The 3D version of this approach was developed by Felten et al.[42] using the analytical form for the corresponding Green's function. Independently, Scrymgeour and Gopalan[43] have used the finite element method to model PFM signals across domain walls.

Here, we analyze PFM contrast for different tip geometries using the Green's function approach originally suggested by Felten et al.[42] and extending recent analysis by Kalinin, Eliseev, and Morozovska.[44] Closed-form expressions for the PFM signal, including relative contributions from individual elements of the piezoelectric constant tensor, elastic properties, and the effects of dielectric anisotropy on the PFM signal, are derived. The solution is developed for transversally isotropic materials corresponding to the case of $c^+$ - $c^-$ domains in tetragonal ferroelectrics in the limit of weak elastic anisotropy and full anisotropic material with weak elastic and dielectric anisotropies.

### III.1. Electric fields distributions

The initial step in calculating the electromechanical response in the decoupled approximation is the determination of the electric field distributions. While for isotropic[45] and transversally isotropic[46] materials, the solution can be obtained using simple image charge method (Fig. 2), the field is significantly more complex in materials with lower symmetry. Here we analyze the case of the full dielectric anisotropy.

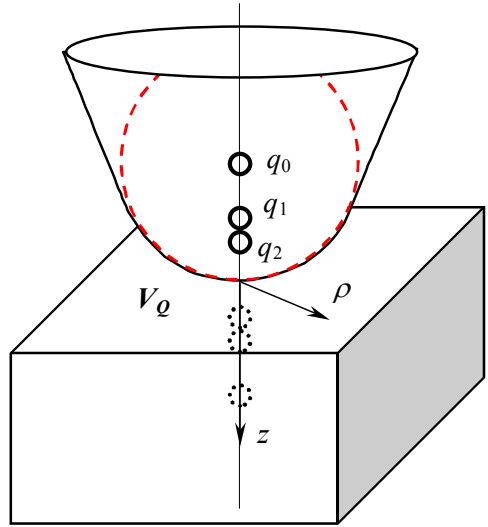

**Fig. 2.** Tip representation using image charge distribution in the PFM experiment.

The potential distribution in the anisotropic half-space with dielectric permittivity, $\varepsilon_{ij}$, induced by a point charge, $Q$, located at the distance, $d$, above the surface can be obtained from the solution of the Laplace equation:

$$\begin{cases} \varepsilon_0 \varepsilon_{ij} \dfrac{\partial^2}{\partial x_i \partial x_j} V(\mathbf{r}) = 0, & z \geq 0 \\ \varepsilon_0 \Delta V_0(\mathbf{r}) = -Q \cdot \delta(z+d)\delta(x)\delta(y), & z < 0 \end{cases} \quad (5a)$$

with the boundary conditions for electric field and potential

$$\varepsilon_{3j} \dfrac{\partial}{\partial x_j} V(z=0) = \dfrac{\partial}{\partial z} V_0(z=0),$$
$$V(z=0) = V_0(z=0) \quad (5b)$$

where $V(\mathbf{r})$ is the potential distribution inside the half-space ($z > 0$) and $V_0(\mathbf{r})$ is the potential distribution in a free space. In a case of general dielectric anisotropy, the electrostatic potential $V(\mathbf{r})$ is found in the Fourier representation (Appendix B) as:

$$V(\mathbf{r}) = \dfrac{Q}{2\pi\varepsilon_0} \int_{-\infty}^{\infty} dk_x \int_{-\infty}^{\infty} dk_y \, \dfrac{\exp(-ik_x x - ik_y y)}{2\pi} \cdot \dfrac{\exp\left(-d\sqrt{k_x^2 + k_y^2} - z\lambda(k_x, k_y)\right)}{\left(\sqrt{k_x^2 + k_y^2} + (i\varepsilon_{31} k_x + i\varepsilon_{32} k_y + \varepsilon_{33}\lambda)\right)} \quad (6a)$$

where

$$\lambda(k_x, k_y) = \dfrac{\sqrt{(\varepsilon_{11}\varepsilon_{33} - \varepsilon_{31}^2)k_x^2 - 2(\varepsilon_{31}\varepsilon_{32} - \varepsilon_{12}\varepsilon_{33})k_x k_y + (\varepsilon_{22}\varepsilon_{33} - \varepsilon_{32}^2)k_y^2} - i(\varepsilon_{31} k_x + \varepsilon_{32} k_y)}{\varepsilon_{33}}$$

(6b)

The square root in Eq. (6b) is real for any real $(k_x, k_y)$ since the dielectric constant tensor, $\varepsilon_{ij}$, is positively defined.

For a transversally isotropic dielectric material ($\varepsilon_{ij} = \varepsilon_{ii}\delta_{ij}$, $\varepsilon_{11} = \varepsilon_{22} \neq \varepsilon_{33}$) Eq. (6) is simplified as

$$V_Q(\rho, z) = \dfrac{Q}{2\pi\varepsilon_0(\kappa+1)} \dfrac{1}{\sqrt{\rho^2 + (z/\gamma + d)^2}}, \quad (7)$$

where $\sqrt{x^2+y^2}=\rho$ and $z$ are the radial and vertical coordinates respectively, $\kappa=\sqrt{\varepsilon_{33}\varepsilon_{11}}$ is the effective dielectric constant, and $\gamma=\sqrt{\varepsilon_{33}/\varepsilon_{11}}$ is the dielectric anisotropy factor.

Numerically, we have shown that Eq. (6) could be used for an orthotropic material symmetry (corresponding e.g. to in-plane $a$ domains in tetragonal ferroelectrics) $\varepsilon_{ij}=\varepsilon_{ii}\delta_{ij}$, $\varepsilon_{11}\neq\varepsilon_{22}\neq\varepsilon_{33}$ after substitution $\kappa=\sqrt{\varepsilon_{33}(\varepsilon_{11}+\varepsilon_{22})/2}$ and $\gamma=\sqrt{2\varepsilon_{33}/(\varepsilon_{11}+\varepsilon_{22})}$ with accuracy of $\frac{|\varepsilon_{11}-\varepsilon_{22}|}{\varepsilon_{11}+\varepsilon_{22}}\times 100\%$. Thus, Eq. (7) can be used to obtain approximate analytical results for materials with dielectric symmetry close to transversely anisotropic. For fully anisotropic materials, a similar approximation is valid if the off-diagonal elements are small.

### III.3. Electromechanical response to a point charge

The decoupling Green's function approach developed by Felten *et al.*[42] is based on the calculation of the (1) electric field for a rigid dielectric ( $d_{ijk}=e_{ijk}=0$ ), (2) stress field $X_{ij}=E_k e_{kij}$ in piezoelectric material and (3) mechanical displacement field for a non-piezoelectric, elastic material. This approach, while not rigorous, significantly simplifies the problem and in particular allows the effective symmetry of the elastic, dielectric, and piezoelectric properties of a material to be varied independently (Appendix A).[44] In particular, we note that the dielectric and particularly elastic properties described by positively defined second- and fourth-rank tensors (invariant with respect to 180° rotation) are necessarily more isotropic than piezoelectric properties described by third-rank tensors (antisymmetric with respect to 180° rotation).

In the framework of this model, the displacement vector $u_i(\mathbf{x})$ at position $\mathbf{x}$ is

$$u_i(\mathbf{x}) = \int_0^\infty d\xi_3 \int_{-\infty}^\infty d\xi_2 \int_{-\infty}^\infty d\xi_1 \frac{\partial G_{ij}(\mathbf{x},\xi)}{\partial \xi_l} E_k(\xi) e_{kjl} \tag{8}$$

where $\xi$ is the coordinate system related to the material, $e_{kjl}$ are the stress piezoelectric tensor coefficients ($e_{kij} = d_{klm} c_{lmij}$, where $d_{klm}$ are the strain piezoelectric coefficients and $c_{lmij}$ are elastic stiffnesses) and the Einstein summation convention is used. $E_k(\xi)$ is the electric field produced by the probe. For most ferroelectric perovskites, the symmetry of the elastic properties can be approximated as cubic (anisotropy of elastic properties is much smaller than those of the dielectric and piezoelectric properties), and therefore, the approximation of elastic isotropy is used. The Green's function for an isotropic, semi-infinite half-plane is[47,48,49,50]

$$G_{ij}(x_1, x_2, x_3 = 0, \xi) = \begin{cases} \frac{1+\nu}{2\pi Y}\left[\frac{\delta_{ij}}{R} + \frac{(x_i - \xi_i)(x_j - \xi_j)}{R^3} + \frac{1-2\nu}{R+\xi_3}\left(\delta_{ij} - \frac{(x_i - \xi_i)(x_j - \xi_j)}{R(R+\xi_3)}\right)\right] & i, j \neq 3 \\ \frac{(1+\nu)(x_i - \xi_i)}{2\pi Y}\left(\frac{-\xi_3}{R^3} \pm \frac{(1-2\nu)}{R(R+\xi_3)}\right) & \begin{array}{l}\text{"+"} \sim j = 1, 2 \text{ and } i = 3 \\ \text{"--"} \sim i = 1, 2 \text{ and } j = 3\end{array} \\ \frac{1+\nu}{2\pi Y}\left(\frac{2(1-\nu)}{R} + \frac{\xi_3^2}{R^3}\right) & i = j = 3 \end{cases}$$

(9)

where $R = \sqrt{(x_1 - \xi_1)^2 + (x_2 - \xi_2)^2 + \xi_3^2}$, $Y$ is Young's modulus, and $\nu$ is the Poisson ratio. Corresponding expressions for transversally isotropic materials are available elsewhere.[51] Finally, for lower material symmetries, closed-form representations for elastic Green's functions are generally unavailable, but approximate solutions can be derived.[48]

### III.3.1. Transversally isotropic dielectric material

For the special case of a transversally isotropic material, the PFM response can be calculated in analytical form assuming weak elastic anisotropy. After lengthy manipulations, Eq. (8) is integrated to yield the normal displacement of the surface ($z = 0$) as[44]

$$u_3(\rho) = \frac{Q}{2\pi\varepsilon_0(1+\kappa)} \frac{1+\nu}{Y} \frac{1}{\sqrt{\rho^2 + d^2}} (e_{31} f_1(\gamma) + e_{15} f_2(\gamma) + e_{33} f_3(\gamma)), \qquad (10)$$

where the functions $f_i(\gamma)$ depend only on the dielectric anisotropy, $\gamma$, and the Poisson ratio, $\nu$, of the material and the Voigt representation, $e_{31} \equiv e_{311}$, $e_{33} \equiv e_{333}$, $e_{15} \equiv e_{113}$, is used when possible. The contributions of different piezoelectric constants are additive and the corresponding functions $f_i(\gamma)$ are

$$f_1(\gamma) = \frac{\gamma}{(1+\gamma)^2} - \frac{(1-2\nu)}{(1+\gamma)} \qquad (11a)$$

$$f_2(\gamma) = -\frac{2\gamma^2}{(1+\gamma)^2} \qquad (11b)$$

$$f_3(\gamma) = -\left( \frac{\gamma}{(1+\gamma)^2} + \frac{(1-2\nu)}{1+\gamma} \right) \qquad (11c)$$

The vertical PFM signal is determined by the surface displacement at the position of the tip, $u_3(0)$. The in-plane components of surface displacement are $(u_1, u_2)$, where $u_1 = x\tilde{u}(\rho, d)$, $u_2 = y\tilde{u}(\rho, d)$ and

$$\tilde{u}(\rho, d) = \frac{Q}{2\pi\varepsilon_0(1+\kappa)} \frac{1+\nu}{Y} \frac{1}{\sqrt{\rho^2+d^2}\left(\sqrt{\rho^2+d^2}+d\right)} (e_{31} g_1(\gamma) + e_{15} g_2(\gamma) + e_{33} g_3(\gamma)). \qquad (12)$$

The functions $g_i(\gamma)$ that determine the contributions of different piezoelectric constants to overall response are

$$g_1(\gamma) = \frac{1}{(1+\gamma)^2} - \frac{(1-2\nu)}{(1+\gamma)}, \tag{13a}$$

$$g_2(\gamma) = -\frac{2\gamma}{(1+\gamma)^2}, \tag{13b}$$

$$g_3(\gamma) = -\frac{1}{(1+\gamma)^2} - \frac{(1-2\nu)}{1+\gamma}. \tag{13c}$$

The polar components of the in-plane displacement vector $u_\rho = (xu_1 + yu_2)/\rho$, $u_\varphi = (-yu_1 + xu_2)/\rho$, are $u_\rho(\rho) = \rho\tilde{u}(\rho,d)$ and $u_\varphi \equiv 0$. The in-plane displacements at contact are $u_1(0) = u_2(0) = 0$, and hence the lateral PFM signals are zero in agreement with the rotational symmetry of the system.

In most materials, the Poisson ratio is $\nu \approx 0.35$. Hence, the functions $f_i(\gamma)$ and $g_i(\gamma)$ that determine the contributions of the piezoelectric constants $e_{in}$ to the overall signal depend primarily on the dielectric anisotropy of the material, $\gamma$. For most ferroelectric oxides $\gamma \approx 0.3 - 1$, while $\gamma \approx 10.5$ and 2.3 for Rochelle salt and triglycine sulfate, respectively.

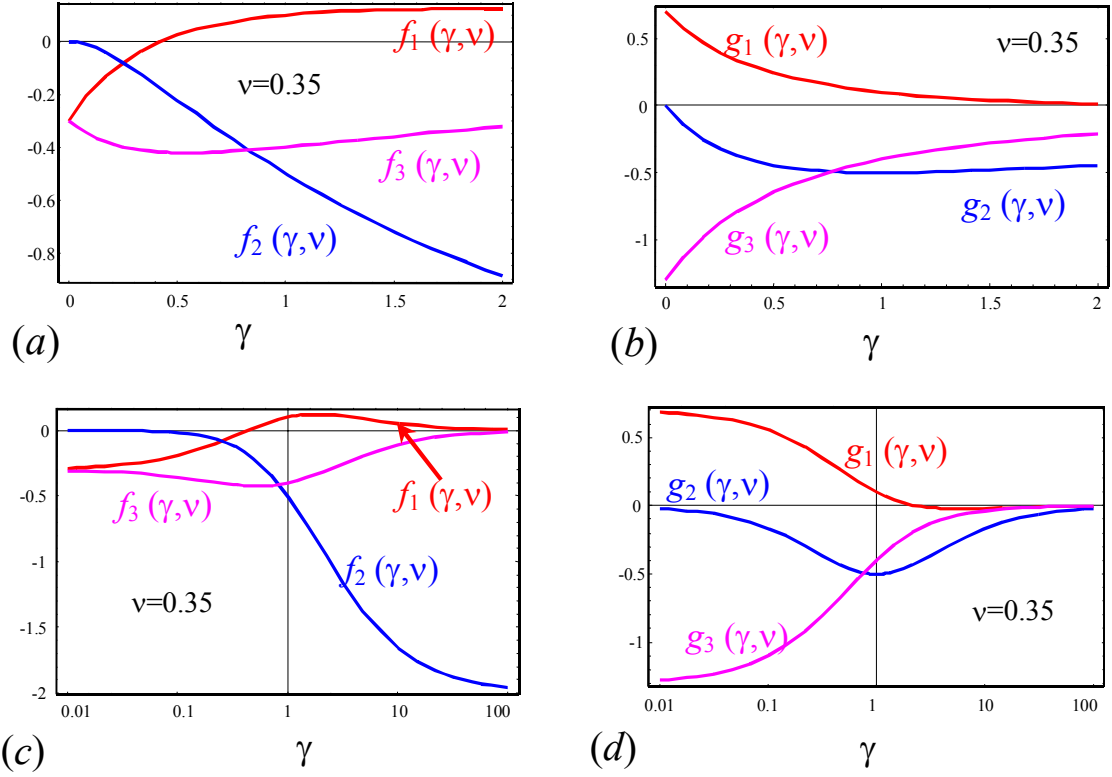

**Fig. 3.** (a,b) Plots of the functions $f_i(\gamma,\nu)$ and $g_i(\gamma,\nu)$ for $\nu = 0.35$ in the (0,1) interval corresponding to most ferroelectric oxides. (c,d) Plots in the logarithmic scale in the ($10^{-2}$, $10^2$) interval illustrating the asymptotic behavior of the dissimilar contributions with dielectric anisotropy.

Shown in Fig. 3 (a,b) are plots of the functions $f_i(\gamma)$ that determine contributions of the piezoelectric constants $e_{nm}$ to the vertical displacement, $u_3$. The function $f_3(\gamma)$ varies relatively weakly with $\gamma$ in the interval $\gamma \in (0,2)$, indicating the weak dependence of the longitudinal contribution, $e_{33}$, to signal. For large dielectric anisotropies: $\gamma \to \infty$ and $f_3(\gamma) \to 0$. Conversely, $f_2(\gamma)$, which determines the contribution of constants $e_{15}$ to the signal, increases rapidly with $\gamma$. Finally, $f_1(\gamma)$ changes sign at $\gamma = 0.4$ and is much smaller than $f_3(\gamma)$, $f_2(\gamma)$. Shown in Fig. 3 (b,d) are the functions $g_i(\gamma)$ that determine contributions of piezoelectric constants $e_{nm}$ to the overall signal $\tilde{u}(\rho,d)$. The functions $g_1(\gamma)$ and $g_3(\gamma)$ decay with $\gamma$, indicating the decreasing contribution of $e_{33}$ and $e_{31}$ to the signal. The signal decreases for low compressibility materials (high v). Conversely, $g_2(\gamma)$, which determines contributions of the constant $e_{15}$ to the signal, increases with $\gamma$ for $\gamma < 1$. For $\gamma > 1$, all $g_i(\gamma)$ decrease.

The point-charge response in Eqs. (10) and (12) can be extended to realistic tip-geometries using an appropriate image charge model, e.g., an image charge series for a spherical tips or a line charge model for conical tips.[52] In particular, from the similarity between Eq. (7) for the potential distribution induced on the surface induced by a point charge and Eq. (10) for surface displacement induced by a point charge, we derive the following principle.

**Response Theorem 1:** For a transversally isotropic piezoelectric solid in an isotropic elastic approximation and an arbitrary point charge distribution in the tip (not necessarily constrained to a single line), the vertical surface displacement is proportional to the surface potential $V_Q(\rho)$ induced by tip charges in the point of contact given by Eq. (7) and

$$u_3(\rho) = V_Q(\rho)\frac{1+\nu}{Y}(e_{31}f_1(\gamma) + e_{15}f_2(\gamma) + e_{33}f_3(\gamma)). \tag{14}$$

This is no longer the case for the in-plane components. In this case, the surface displacement fields are more complex

$$u_1(x,y) = W_{Q1}(x,\rho)\frac{1+\nu}{Y}(e_{31}g_1(\gamma) + e_{15}g_2(\gamma) + e_{33}g_3(\gamma)), \tag{15}$$

$$u_2(x,y) = W_{Q2}(y,\rho)\frac{1+\nu}{Y}(e_{31}g_1(\gamma) + e_{15}g_2(\gamma) + e_{33}g_3(\gamma)), \tag{16}$$

where

$$W_{Q1}(x,\rho) = \frac{1}{2\pi\varepsilon_0(1+\kappa)} \sum_i \frac{Q_i}{\sqrt{\rho^2+d_i^2}} \frac{x}{\left(\sqrt{\rho^2+d_i^2}+d_i\right)} \quad (17a)$$

$$W_{Q2}(y,\rho) = \frac{1}{2\pi\varepsilon_0(1+\kappa)} \sum_i \frac{Q_i}{\sqrt{\rho^2+d_i^2}} \frac{y}{\left(\sqrt{\rho^2+d_i^2}+d_i\right)}. \quad (17b)$$

and $Q_i$ and $d_i$ are the effective charges values and z-coordinates respectively. The Eqs. (16, 17) allow estimating the in-plane response caused by the tip asymmetry.

The dependence of the vertical surface response only on the potential induced on the surface if the tip charges are located on the same line along the surface normal also applies for PFM signals for materials with lower symmetries, as will be analyzed below.

### III.3.2. General piezoelectric anisotropy

One of the key problems in PFM is determining the response for a fully anisotropic material, in which case both normal and in-plane components of the surface displacement can be nonzero. Note, that this case corresponds both to materials with low symmetry (e.g., triclinic and monoclinic) and tetragonal perovskites for the case when the orientation of the crystallographic *c*-axis and the surface normal do not coincide.

In the case of general piezoelectric anisotropy, all elements of the stress piezoelectric tensor, $e_{jlk}$, can be nonzero, necessitating the evaluation of all integrals in Eq. (8). In this case, Eq. (8) is a convolution of a fourth-rank tensor $\left(\partial G_{ij}/\partial \xi_l\right)E_k$ with a third-rank tensor of piezoelectric constants, $e_{kjl}$, which is symmetric on $j$ and $l$. Thus, we symmetrize it on the indexes $j$, $l$ and introduce the symmetrical tensor $W_{ijlk}$. The vector of surface displacement $u_i(\mathbf{x})$ induced by the electric field is

$$u_i(\mathbf{x}) = W_{ijlk}(\mathbf{x})e_{klj}, \tag{18}$$

where the components of tensor $W_{ijlk}(\mathbf{x})$ are

$$W_{ijlk}(\mathbf{x}) = \int_0^\infty d\xi_3 \int_{-\infty}^\infty d\xi_3 \int_{-\infty}^\infty d\xi_1 \frac{1}{2}\left(\frac{\partial G_{ij}(\mathbf{x},\boldsymbol{\xi})}{\partial \xi_l} + \frac{\partial G_{il}(\mathbf{x},\boldsymbol{\xi})}{\partial \xi_j}\right)E_k(\boldsymbol{\xi}). \tag{19}$$

The Eq. (18) thus provides a comprehensive description of PFM image formation mechanism on a fully-anisotropic piezoelectric material in the decoupled approximation. The elements of the response tensor $W_{ijlk}(\mathbf{x})$ can be evaluated numerically or analytically under simplifying assumptions on materials symmetry, as discussed below.

In order to obtain an exact analytical solution for $W_{ijlk}(\mathbf{x})$, all integrations were carried out for $x_3 = 0$. For simplicity and to elucidate the underpinnings of the contrast formation mechanism in PFM of anisotropic materials, the analytical calculations are performed for dielectrically isotropic materials, i.e., $\gamma = 1$. Exact analytical expressions for nonzero displacement field components $W_{ijlk}(x, y, z = 0)$ are given in Appendix C. Note that this approximation, while not rigorous, is a good approximation for a broad class of weakly anisotropic materials such as LiTaO$_3$ and poled ferroelectric ceramics in arbitrary orientations with respect to the surface normal. For stronger dielectric anisotropies, Eq. (19) can be evaluated numerically using exact expressions for the electrostatic potential in Eq. (6)

The electromechanical response in PFM is determined by the components of Eq. (18) evaluated at the origin, $W_{ijlk}(0,0,0)$, where Voigt notation applies on indices $j$ and $l$

$$U_{i\alpha k} \equiv \frac{1}{V_Q(0)}\frac{Y}{1+\nu}\begin{cases} W_{ijlk}(0,0,0), & \alpha = 1...3 \\ 2W_{ijlk}(0,0,0), & \alpha = 4...6 \end{cases}, \tag{20}$$

where $V_Q(0) = \dfrac{Q}{2\pi\varepsilon_0(\kappa+1)}\dfrac{1}{d}$. Note that here we introduced Voigt matrix notations in the piezoelectric tensor without any factors $e_{kjl} = e_{k\alpha}$.[53] In reduced notation, the surface displacement below the tip is:

$$u_i(0) = V_Q(0)\frac{1+\nu}{Y}U_{i\alpha k}e_{k\alpha}, \qquad (21)$$

Eq. (21) can be rewritten in a more transparent form as:

$$u_1(0) = V_Q(0)\frac{1+\nu}{Y}(U_{111}e_{11} + U_{121}e_{12} + U_{131}e_{13} + U_{153}e_{35} + U_{162}e_{26}) \qquad (22a)$$

$$u_2(0) = V_Q(0)\frac{1+\nu}{Y}(U_{121}e_{21} + U_{111}e_{22} + U_{131}e_{23} + U_{153}e_{34} + U_{162}e_{16}) \qquad (22b)$$

$$u_3(0) = V_Q(0)\frac{1+\nu}{Y}(U_{131}(e_{31} + e_{32}) + U_{333}e_{33} + U_{351}(e_{24} + e_{15})) \qquad (22c)$$

The nontrivial elements of tensor $U_{i\alpha k}$ are:

$$U_{111} = -\left(\frac{7 + 6(1-2\nu)}{32}\right), \qquad U_{121} = \left(\frac{3 - 2(1-2\nu)}{32}\right), \qquad (23\text{a-b})$$

$$U_{131} = \left(\frac{1 - 2(1-2\nu)}{8}\right), \qquad U_{162} = -\left(\frac{5 + 2(1-2\nu)}{16}\right), \qquad (23\text{c-d})$$

$$U_{153} = -\frac{3}{4}, \qquad U_{351} = -\frac{1}{4}, \qquad U_{333} = -\left(\frac{1 + 2(1-2\nu)}{4}\right). \qquad (23\text{e-g})$$

Note that for a single point charge the potential on the surface below scales as $V_Q(0) \sim 1/d$. Hence, from Eqs. (7) and (21) we derive the following.

**Response Theorem 2**: For an anisotropic piezoelectric solid in the limit of dielectric and elastic isotropy, the vertical and lateral PFM signals are proportional to the potential on the surface induced by the tip if the tip charges and the point of contact are located on the same line along the surface normal. This condition corresponds to e.g., sphere and line charge models for the tip.

Due to the complex functional form of the displacement functions in Eqs. (C.3a-r, C.4a-i), this is no longer the case when the image charge distribution in the tip and point of contact does not fall on the same line, and analyses of several important cases are given in Section III.3.4. Moreover, the formulae in Appendix C allow the effects of tip assymetry on response to be evaluated.

### III.3.3. Elastically and dielectrically anisotropic materials

The analysis developed in Sections III.3.1 and III.3.2 has yielded analytical expressions for vertical and lateral PFM signals for two important cases, namely a transversally isotropic piezoelectric solid in the limit of elastic isotropy and an anisotropic piezoelectric solid in the limits of elastic and dielectric isotropy. These approximations are well justified for ferroelectric perovskites with a cubic paraelectric phase far from the Curie temperature and with relatively weak piezoelectric coupling, as well as ferroelectric ceramics and polymers poled in the direction normal to the surface. However, in many cases, e.g., materials such as $BaTiO_3$, Rochelle salt, etc., the elastic and particularly the dielectric properties of the material are strongly anisotropic. In these cases, the analysis above becomes semiquantitative and can be improved by numerical evaluation of the integrals in Eq. (8) for an appropriate Green's function for an elastic solid and electrostatic field distribution evaluated by Eqs. (6). However, even in these cases, Eqs. (14)-(16) and (21) still provide general insight into the PFM mechanism because of a much stronger anisotropy in the piezoelectric properties (as compared to the elastic and dielectric properties) that will thus dominate the signal.

### III.4. PFM response for multipole tips

Considered above were simple tip models corresponding to uniform conductive SPM tips. In this and subsequent sections, we consider more complex models, corresponding to SPM tips with shielding or formed by strip lines under different biases (Fig. 4). In these cases, the tip can no longer be represented by a surface of constant potential or image charges of the same sign; rather, more complex potential distributions are required. In addition, these models allow estimation of the contribution of higher-order multipole moments to the surface response even for conventional tips, e.g., to allow consideration of the effects of tip asymmetry. Here, we analyze the PFM signal formation mechanism for tips with complex electrostatics modeled using multipole representations for the tip field.

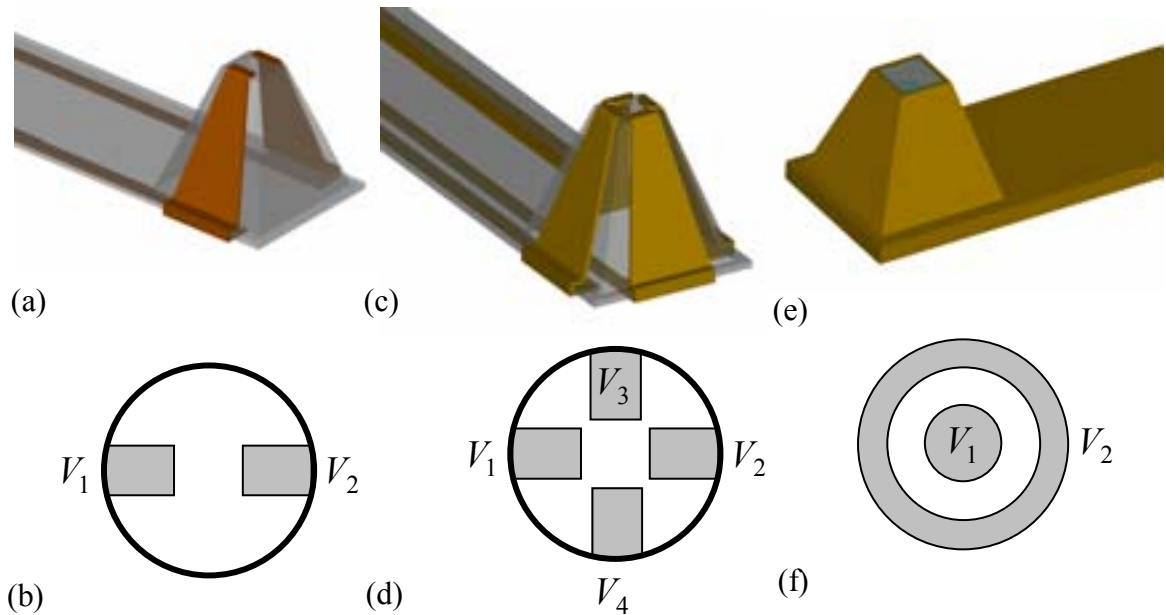

**Fig. 4**. Possible structures for multipole tips. Shown are 3D views and schematics of (a,b) strip line tips, (c,d) quadrupole tip, and (e,f) shielded tip. Strip-line tip can be use to create in-plane dipole field, $V_1 = -V_2 = V_{ac} \cos \omega t$. Quadrupole tip can be used to create quadrupolar electric field, $V_1 = V_2 = V_{ac} \cos \omega t$, $V_3 = V_4 = -V_{ac} \cos \omega t$, or rotating in-plane dipole tip, $V_1 = -V_2 = V_{ac} \cos \omega t$, $V_3 = -V_4 = V_{ac} \sin \omega t$. Shielded tip can be used to localize field and minimize electrostatic contribution to PFM signal, $V_1 = V_{ac} \cos \omega t$, $V_2 = 0$.

### III.4.3. Dipole tip model (vertical)

The simplest example of a multipole AFM tip, in which the electric field has an equipotential line with zero potential at a distance, $a$, from the sample surface. Such potentials can be approximated by two point charges of different signs, $\pm Q_0$, aligned on the surface normal and separated by distance $p$ (Fig. 5).

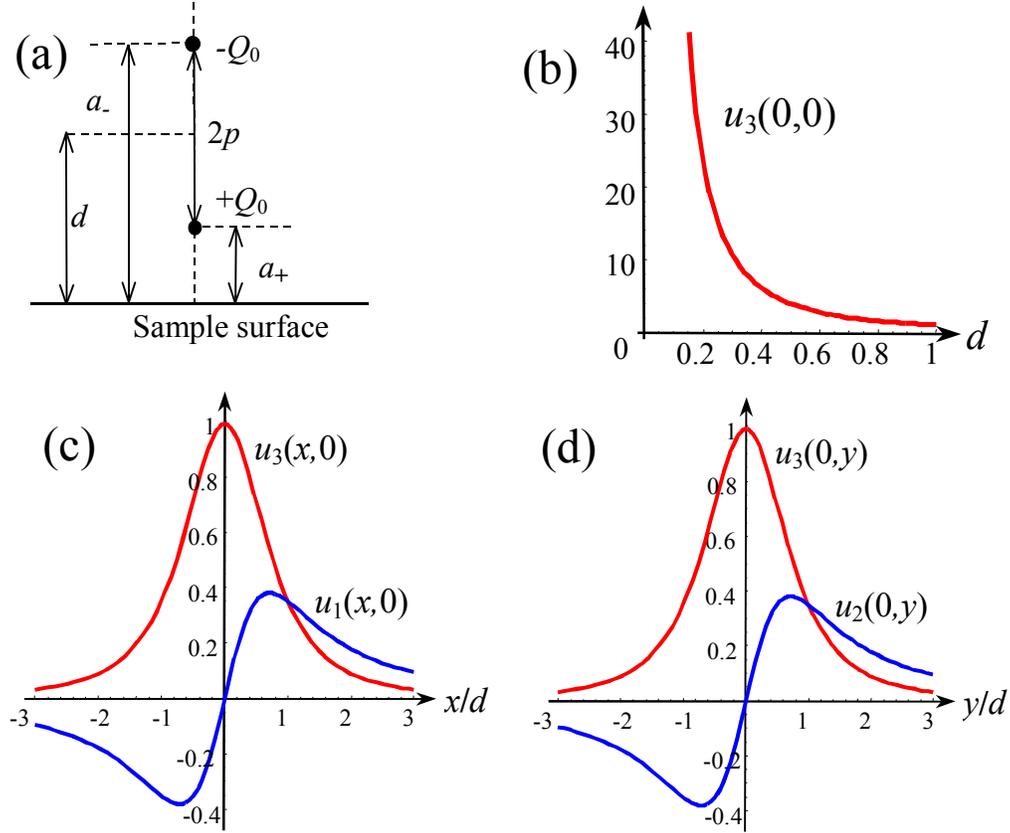

**Fig. 5.** Schematic of the dipole tip (a), (b) vertical displacement $u_3(0,0) \sim V_Q(0)$ vs. the distance of the dipole separation $d$ (arb. units). normalized displacements $u_{1,3}(x,y)$ and $u_{2,3}(x,y)$ in (c) x and (d) y-directions for $p/d = 0.01$.

We introduce the tip dipole moment $P_0 = 2Q_0 p$ when $p \ll d$. The functions $V_Q$ used in Eqs. (14) and (17) are then substituted by

$$V_Q(\rho) \approx \frac{P_0}{2\pi\varepsilon_0(\kappa+1)} \frac{d}{(\rho^2 + d^2)^{3/2}} \qquad (24)$$

The potential on the surface below the tip is

$$V_Q(0) \approx \frac{P_0}{2\pi\varepsilon_0(\kappa+1)d^2} \qquad (25)$$

From the response theorems derived in Sections III.2.1 and III.3.2, the electromechanical response of the surface is proportional to the tip-induced potential, Eqs. (14-16) and (21). This case is thus trivial. The surface displacement fields can be easily calculated from the results in Appendix E. More sophisticated two-charge tip models were discussed by Abplanalp[54] for samples of finite thickness.

### III.4.4. Dipole tip model (horizontal)

The nontrivial behavior can be expected if the electric field below the tip has a large in-plane component. Such fields can be created by standard, pyramidal AFM tips with strip-line type electrodes where one side is biased positively and the other negatively (Fig. 4). Similar tips with four independent electrodes can be used to create dipolar electric field rotating in the surface plane, where the corresponding torsional or flexural component of cantilever response is measured.

The electric field in this case can be modeled using an in-plane dipole with moment $P_0 = 2Q_0 p$ (Fig. 6) formed by charges $\pm Q_0$ located at distances $d$ from the sample surface at distance $p$ from each other. The dipole axis is oriented in $x$-direction.

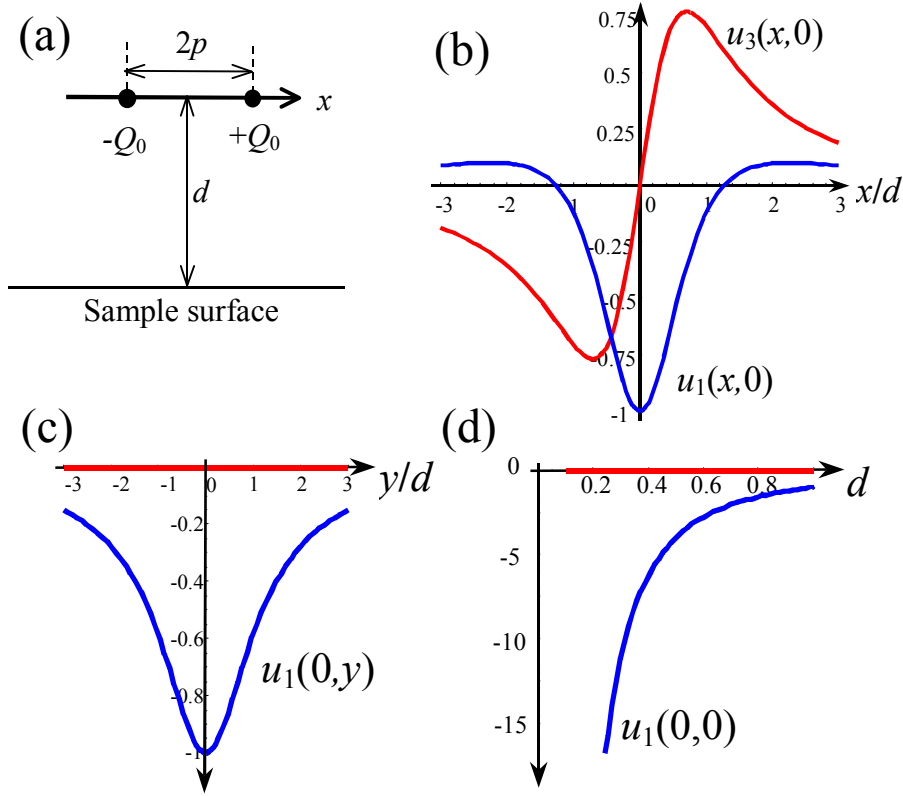

**Fig. 6.** (a) Schematic of the dipole tip. Surface displacement components $u_{1,3}(x,y)$ and $u_1(0,0)$ in (b) x and (c) y directions for $p/d = 0.01$. (d) Surface displacements vs. dipole-surface separation $d$ (arb. units).

For transversally isotropic material, the surface displacement fields can be obtained from Eqs. (14)-(17), where functions $V_Q(\rho)$, $W_{Q1,2}(x,y)$ are substituted by

$$V_Q(x,y) \approx \frac{P_0}{2\pi\varepsilon_0(\kappa+1)} \frac{x}{\left(\sqrt{x^2+y^2+d^2}\right)^3} + O(p^3) \qquad (26a)$$

$$W_{Q1}(x,\rho) \approx -\frac{P_0}{2\pi\varepsilon_0(\kappa+1)} \frac{d(y^2+d^2)+(-x^2+y^2+d^2)\sqrt{x^2+y^2+d^2}}{\left(\sqrt{x^2+y^2+d^2}\right)^3\left(d+\sqrt{x^2+y^2+d^2}\right)^2} + O(p^3) \qquad (26b)$$

$$W_{Q2}(y,\rho) \approx -\frac{P_0}{2\pi\varepsilon_0(\kappa+1)} \frac{xy\left(d+2\sqrt{x^2+y^2+d^2}\right)}{\left(\sqrt{x^2+y^2+d^2}\right)^3\left(d+\sqrt{x^2+y^2+d^2}\right)^2} + O(p^3) \qquad (26c)$$

Assuming that the tip-surface contact corresponds to the center of the dipole, the potential at the contact, derived from Eq. (26a), is zero, $V_Q(0,0) = 0$, and hence, the normal component of the electromechanical response $u_3(0,0)$ is zero. The surface displacement $u_3(x,y)$ is maximal in the points $(x = \pm p, y = 0)$ and has different signs. However, for such a tip, the lateral displacement PFM signal related to in-plane component of surface displacement, $u_1(x,y)$, is now nonzero and can be derived as

$$u_1(0,0) \approx -\frac{P_0}{2\pi\varepsilon_0(\kappa+1)} \frac{1+\nu}{Y} \frac{(e_{31}g_1(\gamma) + e_{15}g_2(\gamma) + e_{33}g_3(\gamma))}{4d^2} \quad (27)$$

The second in-plane component is $u_2(0,0) = 0$, as expected from symmetry considerations. Note that that the distance dependence of the response is $1/d^2$ [see Fig.6 (d)], similar to that of the vertical dipole.

Thus, the use of the dipolar tips potentially allows for probing of different combinations of piezoelectric constants than traditional tips, thus providing additional information about the material. The use of a 4-pole tip (Fig 4b) allows for the generations of a rotating dipole that can potentially provide new operational modes in PFM. Similarly to Eq. (27), the response components of anisotropic materials can be found by using the displacement field components given by Eq. (22).

### III.4.5. Quadrupole tip model

A promising approach for increasing resolution and minimizing the electrostatic force contribution in PFM is based on the use of shielded tips, as shown in Fig. 7. From symmetry considerations, the field produced by such a tip will have monopole and quadrupole components, as considered here. We model the tip with the charge $2Q_0$ located in the point $(0, 0, -d)$ and charges $-Q_0$ are located in cross-points $(\pm p, 0, -d)$. Thus, for $p \ll d$ we can introduce the quadruple moment $K_0 = 2Q_0 p^2$.

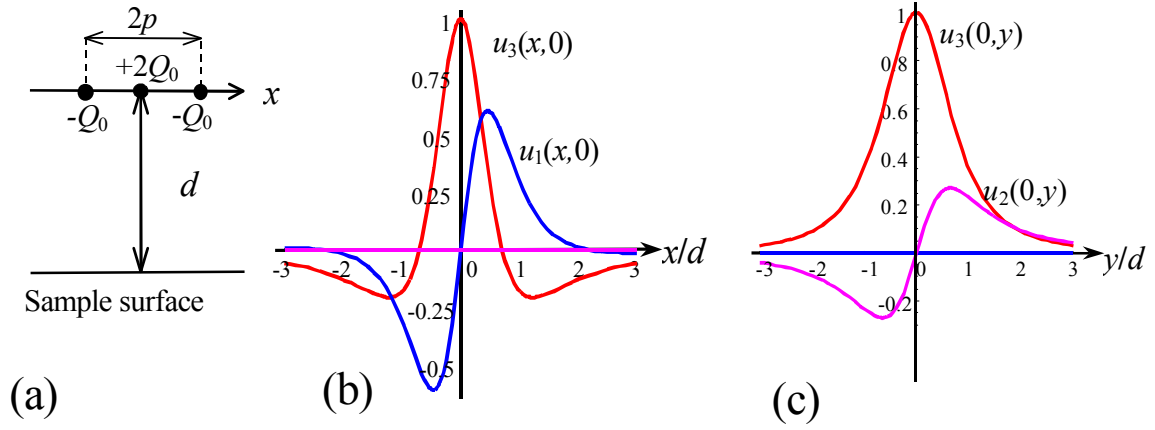

**Fig. 7.** (a) Schematic of the quadrupole tip and normalized displacements in (b) x and (c) y direction for $p/d = 0.01$.

The electromechanical response of the surface for the transversally-isotropic case can be found in a way similar to the dipolar tip from Eqs. (14)-(17), where functions $V_Q(\rho)$, $W_{Q1,2}(x, y)$ are now given by

$$V_Q(x,y) \approx \frac{K_0}{4\pi\varepsilon_0(\kappa+1)} \frac{d^2 - 2x^2 + y^2}{\left(\sqrt{x^2 + y^2 + d^2}\right)^5} \tag{28a}$$

$$W_{Q1}(x,\rho) \approx \frac{K_0}{4\pi\varepsilon_0(\kappa+1)} \frac{x\left(\begin{array}{c} 9d(y^2 + d)\sqrt{x^2 + y^2 + d^2} + 9d^4 + \\ +4d^2 x^2 + 15d^2 y^2 - 2x^4 + 4x^2 y^2 + 6y^4 \end{array}\right)}{\left(\sqrt{x^2 + y^2 + d^2}\right)^5 \left(\sqrt{x^2 + y^2 + d^2} + d\right)^3} \tag{28b}$$

$$W_{Q2}(y,\rho) \approx \frac{K_0}{4\pi\varepsilon_0(\kappa+1)} \frac{y\left(\begin{array}{c} 3d(-2x^2 + y^2 + d)\sqrt{x^2 + y^2 + d^2} + 3d^4 - \\ -6d^2 x^2 + 5d^2 y^2 - 6x^4 - 4x^2 y^2 + 2y^4 \end{array}\right)}{\left(\sqrt{x^2 + y^2 + d^2}\right)^5 \left(\sqrt{x^2 + y^2 + d^2} + d\right)^3} \tag{28c}$$

From Eqs. (28a-c), the lateral displacements below the tip are $u_2(0,0) = u_1(0,0) = 0$, in agreement with the symmetry of the problem. However, the vertical PFM response is now

$$u_3(0,0) \sim \left( \frac{2}{d} - \frac{2}{\sqrt{p^2 + d^2}} \right) \approx \frac{K_0}{d^3} \qquad (29)$$

The approximate Eq. (29) represents the potential on the surface as a function of separation and quadruple moments. As with Eq. (29), the response components of anisotropic materials can be found using the displacement field components given by Eq. (22).

## IV. Discussion

Based on the analysis of PFM, the image formation mechanism in Section III, we discuss the implications for imaging piezoelectric materials. The orientation dependence of the PFM signal and the potential for molecular and crystallographic orientational imaging are discussed in Section IV.1. The distance dependence of the electromechanical and electrostatic contributions to the PFM signal are analyzed in Section IV.2. Finally, the distance dependence of electrostatic and electromechanical contributions to PFM signal and the resolution limits are discussed in Section IV.3.

### IV.1. Orientation dependence

A unique feature of PFM is that in the ideal case, the signal is independent of the tip-surface contact area and is determined solely by material properties. Furthermore, if the contact nonideality leads to a potential drop between the tip and the surface, all of the response components are reduced proportionately. Finally, in the 3D Vector PFM experiment, all three components of the electromechanical response vector can be determined. It has been suggested that these factors allow Vector PFM to be applied to mapping crystallographic and molecular orientation on the nanoscale.[20,55] Briefly, the orientation of a solid body in 3D space is given by three Euler angles (Fig. 8). In PFM, all three components of the displacement vector are measured, which provide three independent equations from which the local Euler angles can be recovered.

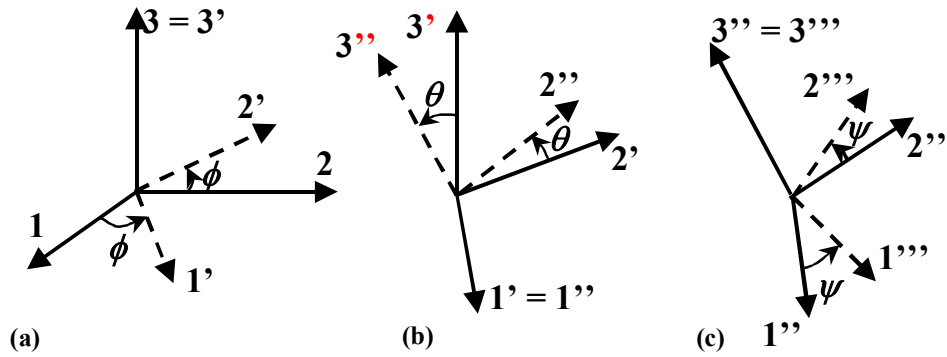

**Fig. 8.** Coordinate transformations for the transition from the crystal to the laboratory coordinate systems. (a) Counterclockwise rotation of $\phi$ about axis 3. (b) Counterclockwise rotation of $\theta$ about axis 1'. (c) Counterclockwise rotation of $\psi$ about axis 3''.

The relationship between the $e_{ijk}$ tensor in the laboratory coordinate system and the $e_{ijk}^0$ tensor in the crystal coordinate system is[56]

$$e_{ijk} = A_{il} A_{jm} A_{kn} e_{lmn}^0, \qquad (30)$$

where $A_{ij}$ is rotation matrix. The displacement components are related to the piezoelectric strain tensor components, as analyzed above. In the particular case of a uniform field, the displacement components detected from vertical and lateral PFM signals are $(u_1, u_2, u_3) = (d_{35}, d_{34}, d_{33})V$, where $d_{33}$ is the longitudinal and $d_{34}, d_{35}$ are the shear components of the strain piezoelectric constant tensor.[20] However, the case of a uniform field rigorously corresponds to systems with a continuous top electrode, which necessarily affects the signal transduction between the surface and the tip and limits the resolution. Moreover, the fabrication of top electrodes for materials such as biopolymers or soft condensed matter systems is not straightforward.

In the PFM geometry, the electric field produced by the tip is strongly non-uniform and the response components are given by Eq. (22). Here, we analyze the applicability of Eqs. (22-23) to the orientational dependence of the PFM data and compare it with the uniform field approximation. As model systems, we consider tetragonal PbTiO$_3$ and trigonal LiTaO$_3$. In the coordinate system oriented along the crystal $c$-axis, the piezoelectric stress $e_{i\alpha}^0$ (or strain $d_{i\alpha}^0$) tensor for these materials are

$$e_{i\alpha}^0 = \begin{pmatrix} 0 & 0 & 0 & 0 & e_{15}^0 & -e_{22}^0 \\ -e_{22}^0 & e_{22}^0 & 0 & e_{15}^0 & 0 & 0 \\ e_{31}^0 & e_{31}^0 & e_{33}^0 & 0 & 0 & 0 \end{pmatrix} \qquad (31)$$

For tetragonal crystals and poled ceramics, $e_{22}^0 \equiv 0$. For cases of an arbitrary crystal orientation, the response components are:

$$e_{33} = (2e_{15}^0 + e_{31}^0)\sin^2\theta\cos\theta + e_{33}^0\cos^3\theta + e_{22}^0\cos\varphi(1-\cos 2\varphi)\sin^3\theta, \qquad (32a)$$

$$e_{34} = \frac{\sin\theta}{2}\begin{pmatrix} -(e_{31}^0 - e_{33}^0 + (2e_{15}^0 + e_{31}^0 - e_{33}^0)\cos 2\theta)\cos\psi + \\ +2e_{22}^0(-\sin\psi\sin 3\varphi + \cos\psi\cos\theta\cos 3\varphi)\sin\theta \end{pmatrix}, \qquad (32b)$$

$$e_{35} = \frac{\sin\theta}{2}\begin{pmatrix} -(e_{31}^0 - e_{33}^0 + (2e_{15}^0 + e_{31}^0 - e_{33}^0)\cos 2\theta)\sin\psi + \\ +2e_{22}^0(\cos\psi\sin 3\varphi + \sin\psi\cos\theta\cos 3\varphi)\sin\theta \end{pmatrix}. \qquad (32c)$$

Similarly, Eqs. (32a-c) apply to strain piezoelectric components if $2e_{15}^0 \rightarrow d_{15}^0$, $e_{31}^0 \rightarrow d_{31}^0$, $e_{33}^0 \rightarrow d_{33}^0$, $2e_{34} \rightarrow d_{34}$, $2e_{35} \rightarrow d_{35}$. Note, that for cases of tetragonal symmetry (i.e.,

BaTiO$_3$) the response is independent of $\varphi$, indicative of the rotational symmetry along the 3-axis.

The dependence of the piezoelectric tensor component $e_{33}$ vs. the orientation of the crystallographic axes with respect to the laboratory coordinate system for LiTaO$_3$ crystal is shown in Fig. 9a. The vertical displacement below the tip vs. the orientation of the crystallographic axes with respect to the laboratory coordinate system for LiTaO$_3$ crystal is shown in Fig. 9b. From the data, the angular dependence of vertical displacement $u_3$, is smoother, more isotropic, and much more convex than the one for $e_{33}$. The maximum value of $u_3$ corresponds to the polar direction $\theta = 0$, whereas $e_{33}$ reaches the maximum for $\theta \neq 0$. From comparisons of Figs. 9a and 9b, it is clear that there are significant differences in the numerical values of the response components. While the directions in which the response is zero and the overall antisymmetric character of the response are the same, as imposed by symmetry consideration, there is significant variability in the numerical values of the response. Even the directions in which the response is maximal can differ. This difference in orientational dependences of $e_{33}(\varphi,\theta)$ and $u_3(\varphi,\theta)$ is due to the fact that according to Eq. (32), the electromechanical response is a sum of contributions due to dissimilar elements of the piezoelectric constant tensor,

$$u_3(\varphi,\theta) \sim (U_{131}(e_{31}(\varphi,\theta)+e_{32}(\varphi,\theta)) + U_{333}e_{33}(\varphi,\theta) + U_{351}(e_{24}(\varphi,\theta)+e_{15}(\varphi,\theta))). \quad (33)$$

Simple numerical estimations prove that the contribution of $U_{333}e_{33}$ into the vertical displacement $u_3(\varphi,\theta)$ dominates for LiTaO$_3$. This fact explains the similarities in the angular dependences. The differences are related the contributions of the other terms in Eq. (33). Also note that the response surfaces for $e_{33}$ and $d_{33}$ are very similar, due to the very weak elastic anisotropy of the material.

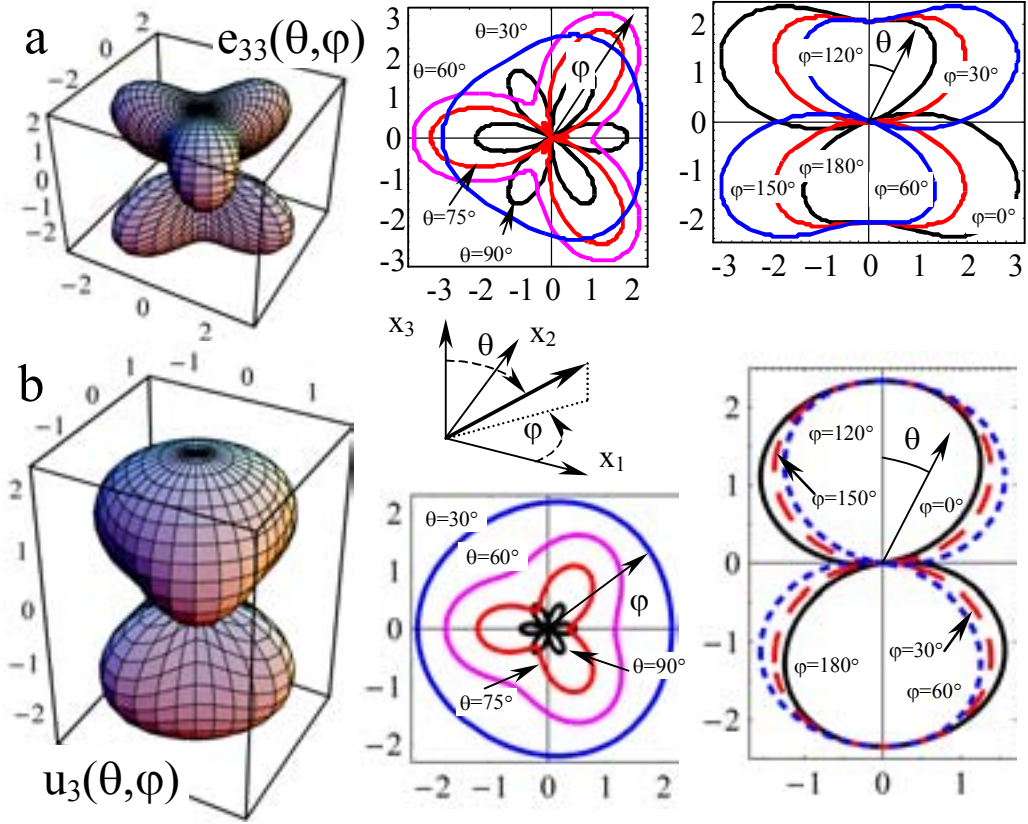

**Fig. 9.** The dependence of (a) piezoelectric tensor component $e_{33}$ and (b) displacement $u_3$ for LiTaO$_3$ on Euler's angles $\varphi$, $\theta$ in the laboratory coordinate system. Components of piezo tensor are the following: $e_{22} = 1.87$, $e_{15} = 2.6$, $e_{31} = -0.1$, $e_{33} = 2.05$ in $C/m^2$. Poisson's ratio is $\nu = 0.25$. Note, that $u_3$ is independent on the Euler angle $\psi$.

The dependence of the piezoelectric tensor component $e_{35}$ vs. the orientation of the crystallographic axes with respect to the laboratory coordinate system for a LiTaO$_3$ crystal is shown in Fig. 10. The horizontal displacement below the tip vs. the orientation of the crystallographic axes with respect to the laboratory coordinate system for a LiTaO$_3$ crystal is shown in Fig. 11.

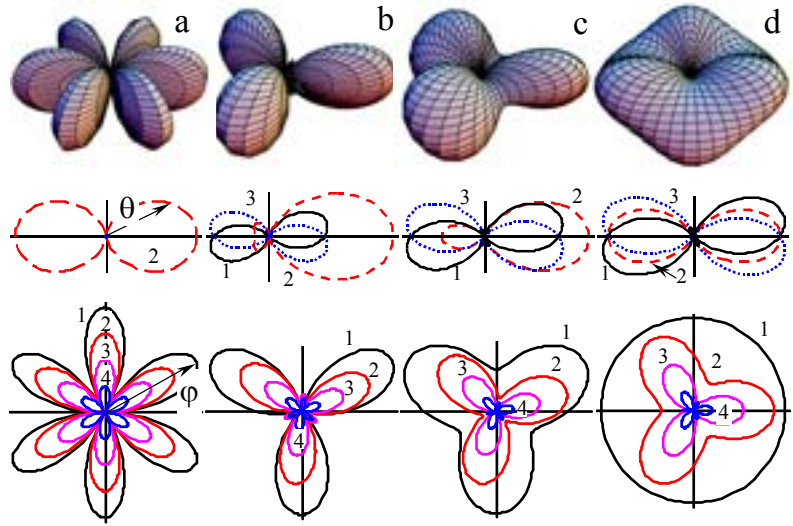

**Fig. 10.** The dependence of piezoelectric tensor component $e_{35}$ on Euler's angles $\varphi$, $\theta$ and $\psi = 0°, 30°, 60°, 90°$ in the laboratory coordinate system (columns a, b, c, d). The upper row represents the 3D view, the middle one is the cross sections at $\varphi = 0°, 30°, 60°$ (curves 1, 2, 3), and the bottom row is the conical sections at $\theta = 90°, 60°, 45°, 30°$ (curves 1, 2, 3, 4).

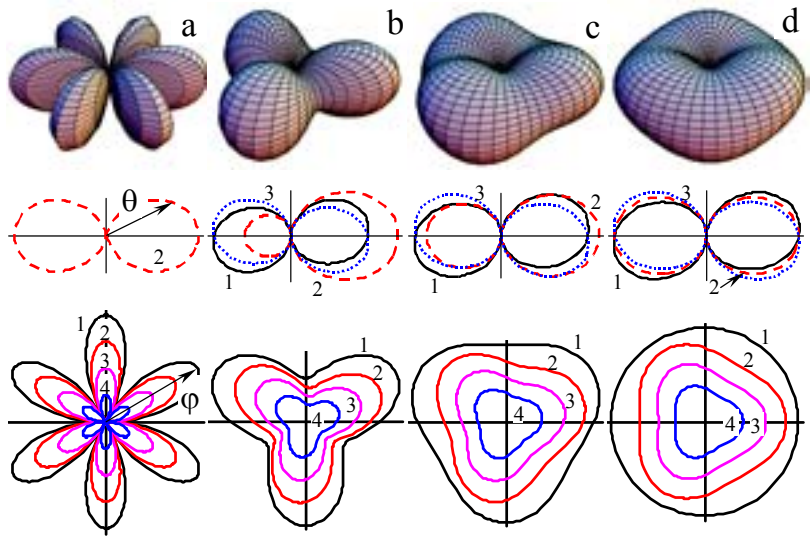

**Fig 11.** The dependence of displacement $u_1$ on Euler's angles $\varphi$, $\theta$ and $\psi = 0°, 30°, 60°, 90°$ in the laboratory coordinate system (columns a, b, c, d). The upper row represents the 3D view, the middle one is the cross sections at $\varphi = 0°, 30°, 60°$ (curves 1, 2, 3), and the bottom row is the conical sections at $\theta = 90°, 60°, 45°, 30°$ (curves 1, 2, 3, 4). Note, that $u_2(\varphi, \theta, \psi) = u_2(\varphi, \theta, 90° - \psi)$.

A common feature of the displacement surfaces shown in Figs. 9–11 is that the $u_1$ angular distribution is smoother, much more symmetric, and convex than the one for $e_{35}$. Similarly to the longitudinal components of the piezoelectric tensors $e_{33}$ and $d_{33}$, the $d_{35}$ surfaces are very similar to $e_{35}$ surfaces.

Vertical displacement below the tip and the piezoelectric tensor components $e_{33}$ and $d_{33}$ vs. the orientation of crystallographic axes with respect to the laboratory coordinate system for a $PbTiO_3$ crystal are shown in Fig. 12 (a-c). Note that $d_{33}$, $e_{33}$, and $u_3$ are independent of the Euler angles $\varphi$ and $\psi$. It is seen that the maximum values of $u_3$ and $d_{33}$ correspond to the polar direction $\theta = 0$, whereas $e_{33}$ reaches the maximum at $\theta \neq 0$.

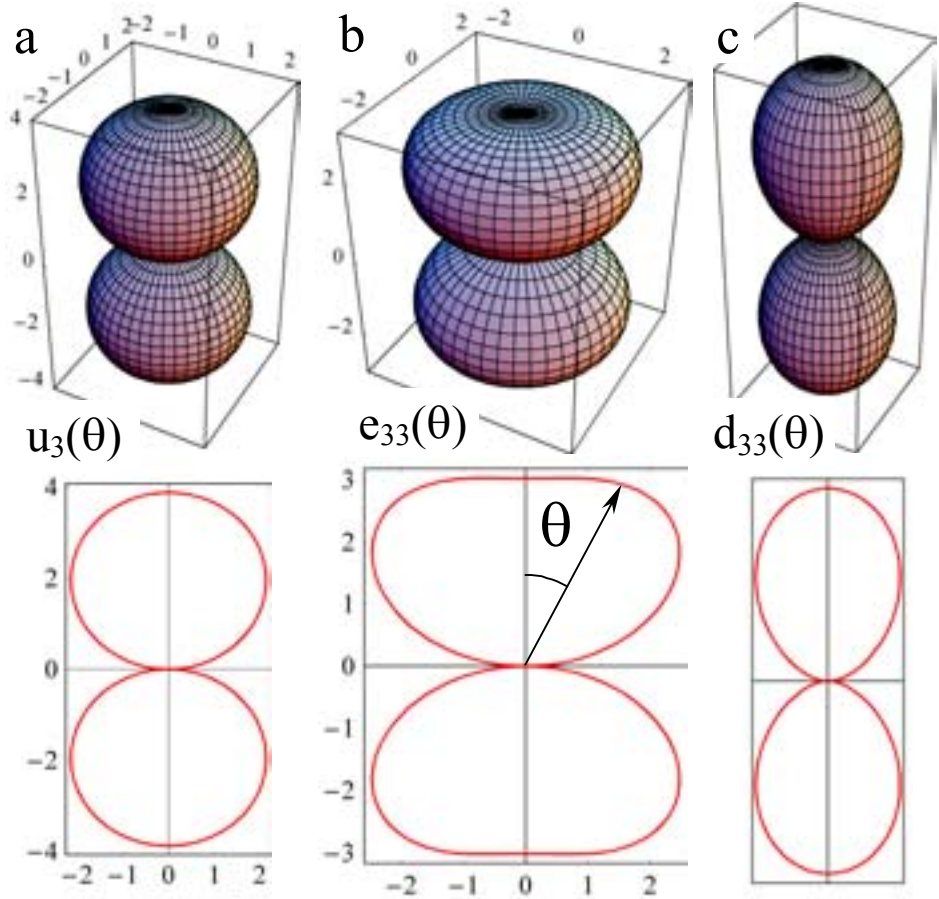

**Fig. 12.** The dependence of (a) displacement $u_3$, (b) piezoelectric tensor component $e_{33}$, and (c) $d_{33}$ for PbTiO$_3$ on Euler's angle θ in the laboratory coordinate system. Components of the piezo tensor in the crystallographic coordinate system are the following: $e_{15}^0 = 4.73$, $e_{31}^0 = -3.12$, $e_{33}^0 = 3.02$ in $C/m^2$. Poisson's ratio is $\nu = 0.3$.

The horizontal displacement component, $u_1$, below the tip, piezoelectric tensor components $e_{35}$ and $d_{35}$ vs. the orientation of the crystallographic axes with respect to the laboratory coordinate system for PbTiO$_3$ is shown in Fig. 13. Note, that the component $u_2$ can be found as $u_2(\varphi, \theta, \psi) = u_1(\varphi, \theta, 90^0 - \psi)$.

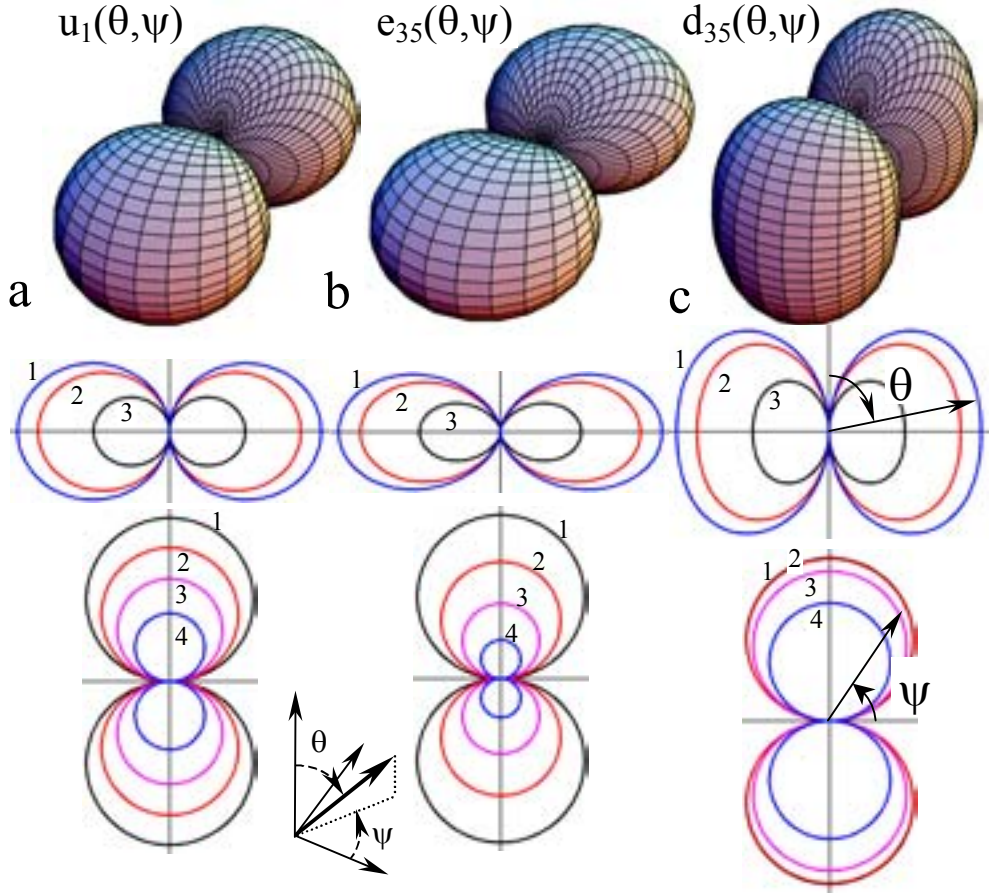

**Fig. 13.** (a) Displacement $u_1$, (b) piezo modulus $e_{35}$, and (c) $d_{35}$ vs. Euler's angles $\psi$, $\theta$ in the laboratory coordinate system for PbTiO$_3$. The upper left-hand side represents the 3D view, the right-hand side is the conical sections at $\theta = 90°, 60°, 45°, 30°$ (curves 1, 2, 3, 4) and the bottom part is the cross sections at $\psi = 30°, 60°, 90°$ (curves 1, 2, 3).

In the analysis above, the dielectric properties of the material were assumed to be close to isotropic and hence the electric field distribution is insensitive to sample orientation. The effect of the orientational dependence of the dielectric properties can be incorporated in a straightforward manner using the analysis in Section III. The displacement in this case is given by a threefold integral:

$$u_i(x,y,z=0) = \int_{-\infty}^{\infty} dk_x \int_{-\infty}^{\infty} dk_y \exp(-ik_x x - ik_y y) \cdot \int_0^{\infty} d\xi\, \widetilde{G}_{ij,l}(k_x,k_y,\xi) \widetilde{E}_k(k_x,k_y,\xi) e_{klj} \qquad (34)$$

Allowing for Eq. (6a), the Fourier representation $\widetilde{E}_k(k_x,k_y,z)$ of the electric field $E_k(\mathbf{r}) = -\dfrac{\partial}{\partial x_k} V(\mathbf{r})$ acquires the form

$$\begin{aligned}
\widetilde{E}_{x,y}(k_x,k_y,\xi) &= ik_{x,y}\frac{Q\exp(-dk - \xi\,\lambda(\mathbf{k}))}{2\pi\varepsilon_0 \left(k + (i\varepsilon_{31} k_x + i\varepsilon_{32} k_y + \varepsilon_{33}\lambda(\mathbf{k}))\right)}, \\
\widetilde{E}_z(k_x,k_y,\xi) &= \lambda(\mathbf{k})\frac{Q\exp(-dk - \xi\,\lambda(\mathbf{k}))}{2\pi\varepsilon_0 \left(k + (i\varepsilon_{31} k_x + i\varepsilon_{32} k_y + \varepsilon_{33}\lambda(\mathbf{k}))\right)}.
\end{aligned} \qquad (35)$$

where $\mathbf{k} = (k_x, k_y)$, $k \equiv \sqrt{k_x^2 + k_y^2}$, and $\lambda(\mathbf{k})$ is given by Eq. (6b). The Fourier representation $\widetilde{G}_{ij,l}(k_x,k_y,\xi)$ of the Green's is given in Appendix A [see (A.10) for details]. The integration on $\xi$ and $k$ in Eq. (34) can be done analytically. Eq. (34) is then reduced to the one-fold integral on the angle $\varphi$ which can be expressed in terms of elliptic integrals. Thus, its calculation is not much more complicated than in the case of a transversely isotropic dielectric media and can be done numerically.

### IV.2. Distance and contact dependence of the PFM signal

One of the key elements in the description of the signal formation mechanism in SPM is the dependence of the signal on tip-surface separation. In the strong indentation limit, corresponding to the classical indentation case, the response is constant, $u_3 = d_3 V_{tip}$ when the tip is in contact, $a > 0$, and zero otherwise (Fig. 14). This behavior is a direct consequence of the boundary conditions employed in a classical indentation problem. However, even when the conductive part of the tip is not in direct contact with the surface (e.g., due to the dielectric gap, oxide layer, etc.), the electric field can partially penetrate into the material, resulting in a nonzero electromechanical response.

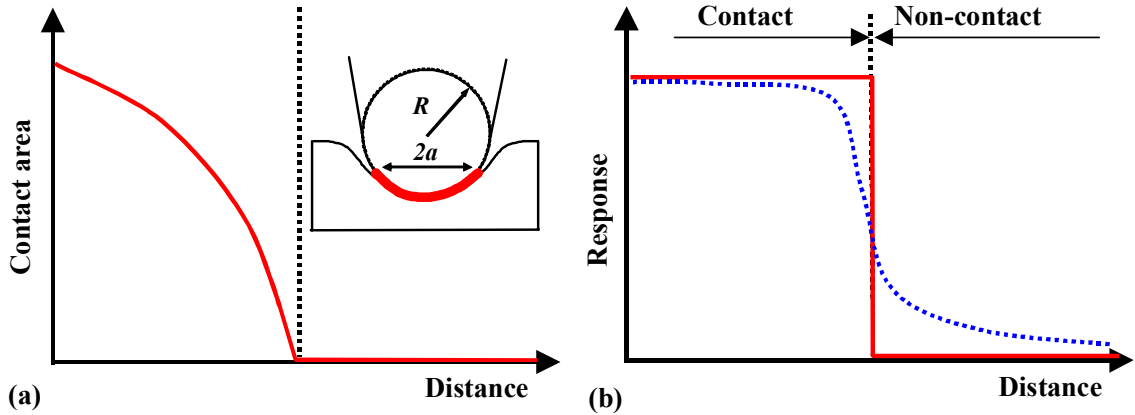

**Fig. 14.** Distance dependence of PFM signal. (a) Dependence of the contact area on tip-surface separation. Inset shows schematics of the contact. (b) Dependence of the electromechanical surface displacement with tip-surface separation in the classical model (solid line) and realistic case. In the contact regime, the response is reduced compared to the classical regime due to contact effects and in the non-contact regime response is nonzero due to partial penetration of the electrostatic field into the surface.

Here, we derive the tip-surface separation dependence for a spherical tip using the response theorem derived in Section III.3. For the spherical part of the tip apex, the tip-induced electric field can be calculated using the image charge model. The solution is rigorous for transversally isotropic material. However, the convergence of image charge series for large dielectric constant mismatch between the ambience and materials requires large number of terms to ensure convergence. For the conical part of the tip, an approximate line-charge model can be used.[52,57] Here, we analyze the electrostatics of tip-surface systems for small separations.

We represent the tip as a charged conductive sphere of radius $R_0$. Its apex is located at distance $\Delta R$ from the sample surface (Fig. 15). Within the framework of this model, all the image charges are on a vertical line and magnitudes and coordinates of the *m*-th image charge are given by

$$q_m = \left(\frac{\kappa-1}{\kappa+1}\right) \cdot \frac{R_0}{2z_0 - d_{m-1}} q_{m-1} \equiv \left(\frac{\kappa-1}{\kappa+1}\right)^m \frac{\text{sh}(\theta)}{\text{sh}((m+1)\theta)} \quad (36a)$$

$$d_m = \frac{R_0^2}{2d - d_{m-1}} \equiv R_0 \frac{\text{sh}(m\theta)}{\text{sh}((m+1)\theta)} \quad (36b)$$

and $q_0 = 1$, $d_0 = 0$, $\text{ch}(\theta) = d/R_0$, $d = R_0 + \Delta R$. Here $Q_0 = 4\pi\varepsilon_0 R_0 U$ is the charge of an isolated tip ($U$ is the voltage applied between the tip and the bottom electrode) and $q_m$ are dimensionless image charges located at distances $d_m$ from the sphere center.

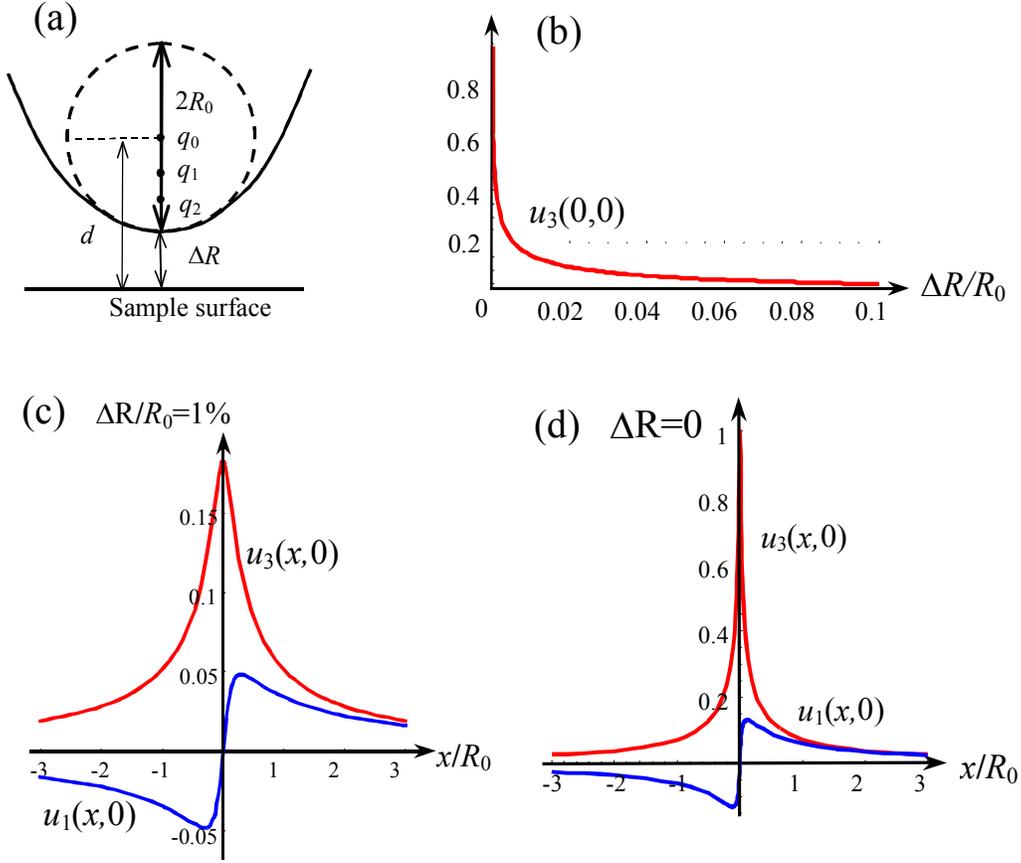

**Fig. 15.** (a) Schematic of the spherical tip, (b) vertical displacement $u_3(0,0) \sim V_Q(0)$ vs. the distance of tip separation $\Delta R/R_0$. (c) x-sections of the normalized displacements for a tip in contact $\Delta R/R_0 = 0$ and for (d) separation $\Delta R/R_0 = 1\%$, all at $p/d = 0.01$.

The components of the surface displacement field are found from Eqs. (14–16), where functions $V_Q$ and $W_{Q1,2}$ are substituted by a series of image charges. While this summation is trivial, these solutions are not easily amenable to analysis. This is especially true of the case for high dielectric constants and small tip-surface separations, where the convergence is slow and summation requires a large number of terms. Here we derive a general Pade approximation for potential on the surface induced by a sphere obtained from the exact series

Eqs. (24a,b) in the case when there is a small gap between the tip apex and the sample surface. For $\Delta R \ll R_0$ and $\kappa \gg 1$ the corresponding Pade approximation has the form

$$V_Q(\rho) \approx \frac{2R_0 U}{\sqrt{\rho^2 + d^2}} \left( \frac{1}{\kappa(1-\exp(-\mu\varphi(\rho))) + (1+\exp(-\mu\varphi(\rho)))} + \frac{\exp(\theta)}{\kappa - 1} \ln\left(\frac{\kappa+1}{\kappa+1-(\kappa-1)\exp(-\theta)}\right) \frac{\rho}{\sqrt{\rho^2 + d^2}} \right) \quad (37)$$

where $\cosh\varphi(\rho) = \sqrt{\rho^2 + d^2}/R_0$, $\cosh(\theta) = d/R_0$, and $0 < \mu \leq 1$ is a fitting parameter. The approximation Eq. (37) and the exact solution are compared in Appendix E. From Eq. (37), the potential on the surface below the tip is

$$V_Q(0) \approx 2R_0 U \frac{1}{d} f\left(\frac{d}{R_0}\right), \quad (38)$$

where

$$f(x) = \frac{1}{\kappa(1-\exp(-\mu\operatorname{arccosh} x)) + (1+\exp(-\mu\operatorname{arccosh} x))}. \quad (39)$$

In contact, $f(1) = 0.5$, and thus the surface potential is equal to tip potential, $V_Q(0) = U$ at $R_0 = d$, recovering correct behavior.

As shown in Sections III.3.1 and III.3.2, for piezoelectric materials of an arbitrary symmetry in the weak elastic and dielectric anisotropy limits and transversally isotropic materials in the weak elastic anisotropy limit, all components of the PFM response are proportional to potential on the surface induced by the tip, if the tip can be represented by the image charged on the normal (Response theorems 1,2). Hence, Eqs. (37–39) describe the distance dependence of the PFM signal for a spherical tip when the tip is above the surface.

## IV.3. Electrostatic vs. electromechanical contributions to PFM signal

One of the key factors that affect PFM imaging is the effect of electrostatic forces. Electrostatic interactions result in a linear (in dc) tip bias contribution to the PFM signal, which does not allow unambiguous separation of the bias-independent (for piezoelectric materials) piezoelectric signal. The electrostatic signal contains two primary contributions: electrostatic forces acting on the tip that results in a second component of surface deformation, and distributed electrostatic force acting on the cantilever that result in additional nonlocal contribution due to flexural vibrations of the cantilever. From the equivalent mechanical model shown in Fig. 16, the PFM signal in the low-frequency limit can be written as

$$PR = \alpha_a(h_0) d_3 \frac{k_1}{k_1 + k} + \frac{C'_{sphere} + C'_{cone}}{k_1 + k}(V_{dc} - V_s) + \frac{C'_{cant}}{24k}(V_{dc} - V_{av}), \qquad (40)$$

The electromechanical response is determined by the effective electromechanical response of the material, $d_3$, and the ratio of the ac tip potential to the ac surface potential of the ferroelectric in ambient (i.e., the potential drop in the tip-surface gap of thickness, $h_0$), $\alpha_a(h_0)$. The electrostatic force contribution is governed by capacitance z-gradients due to the spherical, $C'_{sphere}$, and conical, $C'_{cone}$, parts of the tip and the cantilever, $C'_{cant}$, respectively. The bias dependence of the electrostatic contribution is controlled by the dc potential offset of the tip bias, $V_{dc}$, the domain-dependent surface potential below the tip, $V_s$, and surface potential averaged over the cantilever length, is $V_{av}$. Signal transduction between the surface displacements and the electrostatic forces and flexural cantilever deflection is governed by the spring constant of the tip-surface junction, $k_1$, and the spring constant of the cantilever, $k$. Note the difference in the dependence of the electromechanical and electrostatic contributions on the cantilever and tip-surface spring constants. The factor of 24 in the nonlocal cantilever contribution and the dependence solely on the cantilever spring constant originates from the cantilever modes (buckling in which tip position is constant vs. flexural mode with tip displacement) induced by electrostatic forces acting on the cantilever.[21]

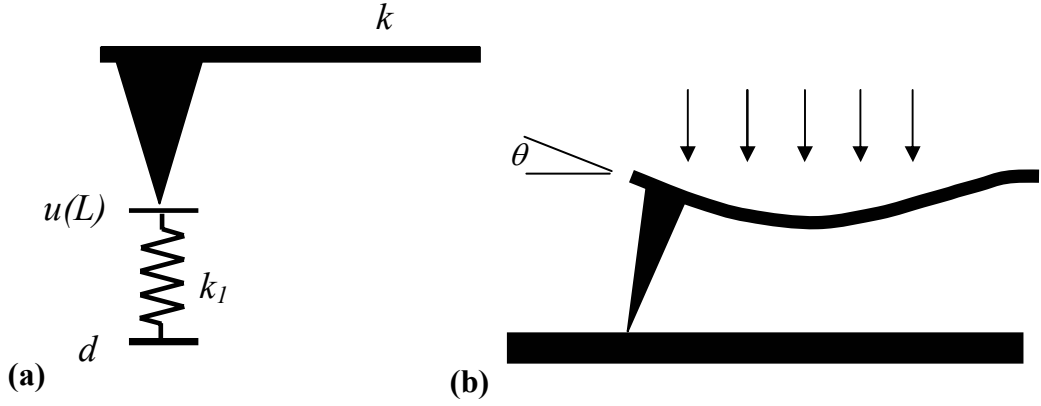

**Fig. 16.** (a) The equivalent circuit for tip-surface junction is formed by two springs with cantilever constant $k_1$ (tip-surface junction) and $k$ (cantilever), the ration of which determine signal transduction from surface to the cantilever. (b) Distributed electrostatic forces acting on the cantilever result in buckling oscillations that couple to vertical deflection signal. The buckling amplitude is determined primarily by the spring constant of the cantilever.

Here, we consider local effects on the PFM signal relevant to high-resolution imaging. The signal due to cantilever-surface interactions is nonlocal, and hence does not contribute to nanoscale contrast. A similar argument applies to the signal produced by the conical part of the tip located at significant distance from the surface. Furthermore, the conical and cantilever contributions can be reduced by using shielded probes. Hence, we consider the effects due to the spherical part of the tip. In this case, Eq. (40) can be simplified as

$$PR = \alpha_a(h)d_3 \frac{k_1}{k_1+k} + \frac{C'_{sphere}}{k_1+k}(V_{dc}-V_s), \qquad (41)$$

To analyze the distance dependence of the PFM signal, we use a simple Hertzian approximation for the tip-surface contact. The relationship between the indentation depth, $h$, tip radius of curvature, $R_0$, and load, $P$, is[58]

$$h = \left(\frac{3P}{4E^*}\right)^{\frac{2}{3}} R_0^{-\frac{1}{3}} \qquad (42)$$

where $E^*$ is the effective Young's modulus of the tip-surface system. The contact radius, $a$, is related to the indentation depth as $a = \sqrt{hR_0}$. The contact stiffness is given by $k_1 = (\partial h/\partial P)^{-1}$, and from Eq. (42), $k_1 = 2aE^*$, or

$$k_1 = 2E^* \sqrt{hR_0} = \left(6PE^{*2}R_0\right)^{\frac{1}{3}} \qquad (43)$$

Shown in Fig. 17a is the distance dependence of the electrostatic and electromechanical contributions to the PFM signal calculated for $R = 50$ nm, $V_{dc} = 0.1$ V, $E^* = 100$ GPa, $d_3 = 50$ pm/V, and $k = 1$ N/m and 40 N/m. The distance dependence of the capacitive tip-surface forces and tip-induced surface potential was calculated from Eqs. (36 a,b) for $h > 0$, neglecting the changes in the sphere area due to contact. Note that the electrostatic contribution decreases rapidly with penetration depth due to changes in the tip-surface stiffness constant. In comparison, shown in Fig. 17 (b) is the fraction of the electromechanical contribution depending on penetration depth. The immediate consequence of Fig. 17 is that quantitative probing of the electromechanical response requires using cantilevers with large spring constants in order to minimize nonlocal electrostatic contributions and at large indentation forces in order to maximize the spring constant of the tip-surface junction. These requirements have long been established as guidelines for quantitative imaging.[21,22,23] Note that in the Hertzian model, the electromechanical signal dominates for a penetration depth of ~1 A, corresponding to a contact radii on the order of ~2 nm for $R = 50$ nm, imposing a limit on the spatial resolution of the technique.

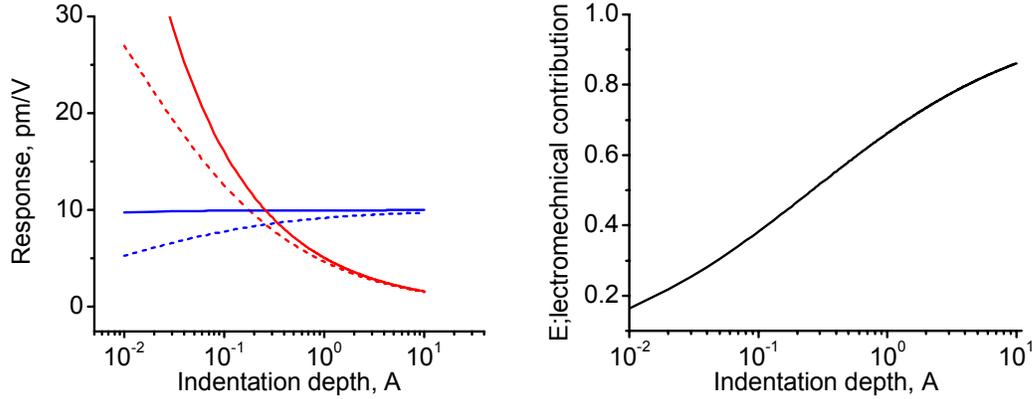

**Fig. 17**. (a) Indentation depth dependence for electromechnical (blue) and electrostatic (red) contributions for a cantilever with k = 1 N/m (solid) and k = 40 N/m (dash). (b) Indentation depth dependence of the fraction of the electromechanical contribution.

The analysis becomes more complicated if adhesive effects are taken into account. In this case, the contact mechanics are described by the Johnson-Kendall-Roberts model.[38,39] In this case, the contact radius is

$$a^3 = \frac{R_0}{E^*}\left\{P + 3\sigma\pi R_0 + \sqrt{6\sigma\pi R_0 P + (3\sigma\pi R_0)^2}\right\} \quad (43)$$

where $\sigma$ is the work of adhesion, $P$ is the load and indentation depth is

$$h = \frac{a^2}{R_0}\left[1 - \frac{2}{3}\left(\frac{r_0}{a}\right)^{3/2}\right] \quad (44)$$

where $r_0^2 = 6\sigma\pi R^2/E^*$ is contact radius at zero force. Shown in Fig. 18 (a) are force vs. indentation depth curves calculated for σ = 0 (Hertzian), $10^{-3}$, $10^{-2}$, $10^{-1}$, and 1 J/m². Shown in Fig. 18 (b,c) are corresponding contact stiffnesses. Note that adhesive contact results in rapid change of contact stiffness from 0 to the value corresponding to contact, resulting in a well-defined boundary between free and bound cantilevers. Finally, shown in Fig. 18 (d) is the force dependence of the contact area. Even for a small work of adhesion, the contact radii at zero force are relatively large, on the order of nanometers. Corresponding contact stiffnesses are of the order of 100–1000 N/m, well above the typical spring constant of the cantilever.

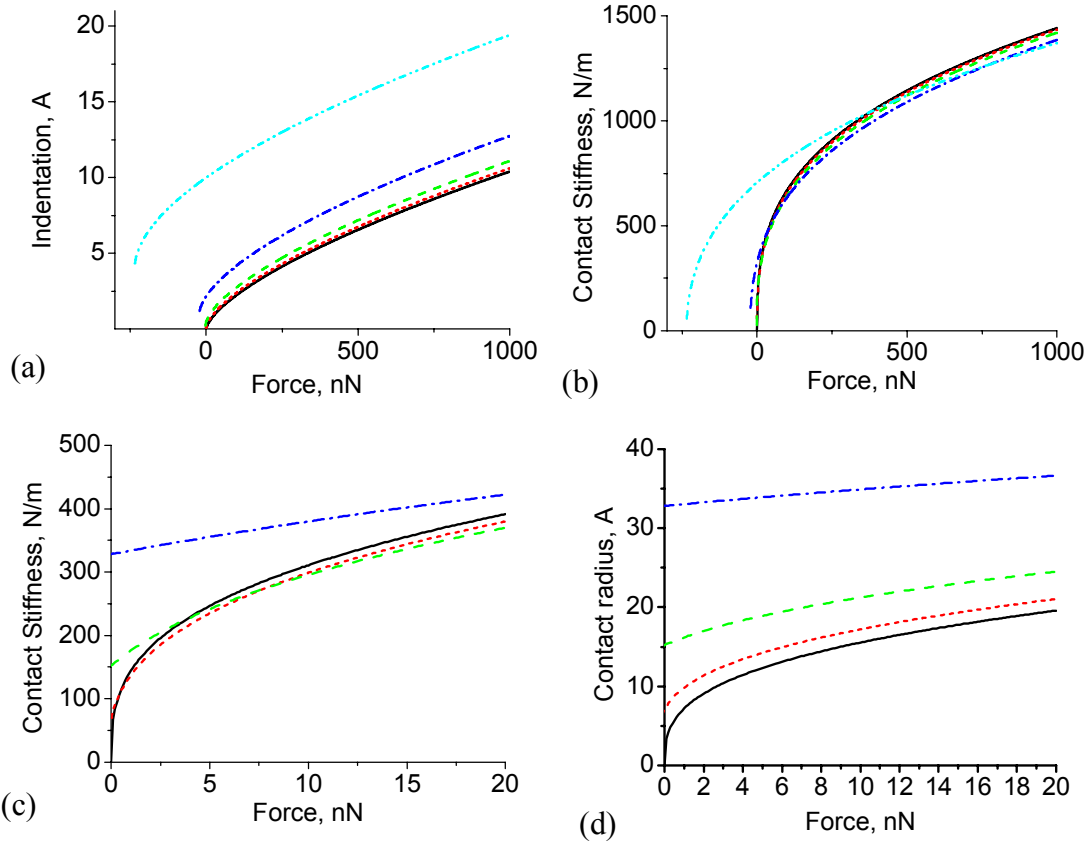

**Fig. 18.** (a) Indentation force-distance dependence in the JKR model. (b,c) force dependence of indentation stiffness. (d) Force dependence of contact radius. Shown are curves for $\sigma = 0$ (solid), $10^{-3}$ (dot), $10^{-2}$ (dash), $10^{-1}$ (dash-dot), and 1 J/m$^2$ (dash-dot-dot).

In both Hertzian and JKR models, the transition to the predominantly electromechanical contrast occurs for contact areas larger than a certain critical value. From Eq. (40), this condition can be generalized for materials with arbitrary properties as

$$a > a^* = \frac{C'_{sphere}(V_{dc} - V_s)}{2\alpha(h)d_3 E^*} \qquad (45)$$

where $a^*$ is the critical contact radius corresponding to equality of the electrostatic and electromechanical contributions to the signal. For classical ferroelectric materials (100 GPa, 50 pm/V), this conditions becomes $a > a^* = 5(V_{dc} - V_s)$ Å. For soft materials, (10 GPa, 5 pm/V), $a > a^* = 50(V_{dc} - V_s)$ nm. Provided that the electrostatic contribution to the signal is minimized, the resolution of PFM on hard materials can potentially achieve sub-nanometer. For soft systems, the signal is likely to represent the convolution of electrostatic and electromechanical signals.

### V. Summary

The image formation mechanism in SPM is analyzed in terms of the tensorial nature of a measured signal and its dependence on the contact radius. It is shown that the PFM signal is only weakly dependent on contact area, distinguishing this technique from AFM and STM. The applicability of linear decoupling approximation for PFM signal is analyzed. The analytical expressions for the PFM signal for different multipole tip models were obtained for transversally isotropic and anisotropic piezoelectric materials as linear combinations of piezoelectric constants. The general formulae for PFM signal for materials with full piezoelectric and dielectric anisotropies are derived. This provides the description of vertical and lateral PFM signal formation mechanism for anisotropic materials.

The dependence of the PFM signal on crystallographic orientation has been analyzed to provide a framework for deconvolution of local crystallographic or molecular orientation from the three components of electromechanical response vector measured by PFM. The effect of contact geometry and imaging conditions on relative contributions of electrostatic forces and piezoelectric response is determined. It is shown that in ambient conditions, the jump of contact stiffness from zero (free cantilever) to large values in contact as limited by the capillary and adhesive forces clearly separates Kelvin Probe Force Microscopy and PFM. This is no longer the case in liquid environment, where the contact stiffness of tip-surface system can be varied continuously as a function of tip-surface separation. The limitations imposed by electrostatic forces on PFM resolution are analyzed.

## Appendix A. Decoupling approximation and Fourier representation for Green's function.

For a linear piezoelectric material, the relationship between strain $U_{ij}$, displacement $D_i$, stress $X_{kl}$, and electric field $E_m$ is

$$U_{ij} = s_{ijkl} X_{kl} + E_m d_{mij}, \tag{A.1}$$

$$D_i = d_{ijk} X_{jk} + \varepsilon_{im} E_m. \tag{A.2}$$

The applicability of decoupling approximation can be established as follows. Eq. (A.2) can be rewritten as $E_k = \varepsilon_{ki}^{-1} D_i - \varepsilon_{ki}^{-1} d_{ijl} X_{jl}$, where the second term is a contribution to the electric field due to deformation. Then from Eq. (A.1) strain components are $U_{ij} = \left(1 - s_{ijkl}^{-1} \varepsilon_{mp}^{-1} d_{plk} d_{mij}\right) s_{ijkl} X_{kl} + \varepsilon_{mp}^{-1} d_{mij} D_p$ where the first term gives the change in compliance due to piezoelectric coupling. Hence, the applicability of the decoupling approximation is determined by the smallness of the dimensionless term $s_{ijkl}^{-1} \varepsilon_{mp}^{-1} d_{plk} d_{mrs}$, i.e., the square of the dimensionless electromechanical coupling coefficients tensor. Moreover, the smaller the electromechanical coupling coefficients, the larger the relative contribution of the dielectric term in Eq. (A.2), justifying the use of a rigid dielectric approximation for calculating the electric field in the material. This analysis is corroborated by the exact solution for the transversally isotropic case.[36,37]

To determine the displacement components in the decoupled approximation, we proceed as follows. Multiplying Eq. (A.1) by the tensor of elastic stiffnesses $c_{ijpq}$ ($\hat{c} \cdot \hat{s} = \hat{I}$) yields

$$c_{ijpq} U_{ij} = c_{ijpq} s_{ijkl} X_{kl} + E_m d_{mij} c_{ijpq}, \tag{A.3}$$

hence

$$c_{ijpq}U_{ij} - E_m d_{mij} c_{ijpq} = X_{pq}, \tag{A.4}$$

and

$$X_{pq} = c_{pqij}U_{ij} - E_m e_{mpq}, \tag{A.5}$$

where $e_{mpq} = d_{mij} c_{ijpq}$ is the piezoelectric stress coefficients. The tensor $X_{pq}$ must satisfy the equilibrium conditions $\partial X_{pq}/\partial x_p = 0$, thus $c_{pqij} \partial U_{ij}/\partial x_p = \partial(E_m e_{mpq})/\partial x_p$. Therefore, the force density is $F_k = -\partial(E_i e_{ijk})/\partial x_j$. The pressure acting on the sample surface $z = 0$ is $P_k = -E_i e_{i3k}$. The displacement is thus:

$$u_i(\mathbf{x}) = -\int_0^\infty \int_{-\infty}^\infty \int_{-\infty}^\infty G_{ij}(\mathbf{x}, \xi_1, \xi_2, \xi_3) \frac{\partial E_k(\xi_1, \xi_2, \xi_3)}{\partial \xi_l} e_{klj} d\xi_1 d\xi_2 d\xi_3 -$$
$$- \int_{-\infty}^\infty \int_{-\infty}^\infty G_{ij}(\mathbf{x}, \xi_1, \xi_2, \xi_3 = 0) E_k(\xi_1, \xi_2, \xi_3 = 0) e_{k3j} d\xi_1 d\xi_2 \tag{A.6}$$

Integration by parts of Eq. (A.6) leads to the following expression:

$$u_i(\mathbf{x}) = \int_0^\infty \int_{-\infty}^\infty \int_{-\infty}^\infty \frac{\partial G_{ij}(\mathbf{x}, \xi_1, \xi_2, \xi_3)}{\partial \xi_l} E_k(\xi_1, \xi_2, \xi_3) e_{klj} d\xi_1 d\xi_2 d\xi_3 \tag{A.7}$$

Here, Green's tensor component $G_{ij}(\mathbf{x}, \boldsymbol{\xi})$ determines the displacement $u_i(\mathbf{x})$ at the point $\mathbf{x}$ under the point force $\mathbf{F}$ component "$j$" applied at the point $\boldsymbol{\xi}$. It is defined by the relation $u_i = G_{ij} F_j$. Relevant for PFM are components of displacement at the surface ($x_3 = 0$):[48, 49, 50]

$$G_{ij}(x_1,x_2,x_3=0,\xi)=\begin{cases}\dfrac{1+\nu}{2\pi Y}\left[\dfrac{\delta_{ij}}{R}+\dfrac{(x_i-\xi_i)(x_j-\xi_j)}{R^3}+\dfrac{1-2\nu}{R+\xi_3}\left(\delta_{ij}-\dfrac{(x_i-\xi_i)(x_j-\xi_j)}{R(R+\xi_3)}\right)\right] & i,j\neq 3\\[2mm]\dfrac{(1+\nu)(x_i-\xi_i)}{2\pi Y}\left(\dfrac{-\xi_3}{R^3}-\dfrac{(1-2\nu)}{R(R+\xi_3)}\right) & i=1,2\ \text{and}\ j=3\\[2mm]\dfrac{(1+\nu)(x_j-\xi_j)}{2\pi Y}\left(\dfrac{-\xi_3}{R^3}+\dfrac{(1-2\nu)}{R(R+\xi_3)}\right) & j=1,2\ \text{and}\ i=3\\[2mm]\dfrac{1+\nu}{2\pi Y}\left(\dfrac{2(1-\nu)}{R}+\dfrac{\xi_3^{\,2}}{R^3}\right) & i=j=3\end{cases}$$

(A.8)

where $R=\sqrt{(x_1-\xi_1)^2+(x_2-\xi_2)^2+\xi_3^2}$. The tensor $G_{ij}$ is nonsymmetrical, since $G_{13}(\mathbf{x};\boldsymbol{\xi})\neq G_{31}(\mathbf{x};\boldsymbol{\xi})$ and $G_{23}(\mathbf{x};\boldsymbol{\xi})\neq G_{32}(\mathbf{x};\boldsymbol{\xi})$, while $G_{12}(\mathbf{x};\boldsymbol{\xi})=G_{21}(\mathbf{x};\boldsymbol{\xi})$.

Since the electric field distribution for the general case of a material with arbitrary dielectric anisotropy can be derived only in the form of Fourier integrals (see Appendix B), calculations of the displacement field require Fourier representation for the Green's function. The components (A.8) depend only on the differences $x_1-\xi_1$ and $x_2-\xi_2$, hence 2D Fourier transform can be introduced as

$$G_{ij}(x_1,x_2,x_3=0,\xi)=\dfrac{1}{2\pi}\int_{-\infty}^{\infty}dk_1\int_{-\infty}^{\infty}dk_2\,\exp(-ik_1(x_1-\xi_1)-ik_2(x_2-\xi_2))\cdot\widetilde{G}_{ij}(k_1,k_2,\xi_3)$$

(A.9)

where the Green's function components in Fourier representation are:[49]

$$\widetilde{G}_{ij}(k_1,k_2,\xi)=\dfrac{1+\nu}{2\pi Y}\dfrac{\exp(-\xi k)}{k}\left(2\delta_{ij}-\dfrac{k_ik_j}{k^2}(\xi k+2\nu)\right),\quad i,j\neq 3 \qquad\text{(A.10a)}$$

$$\widetilde{G}_{i3}(k_1,k_2,\xi)=-\dfrac{1+\nu}{2\pi Y}\cdot\dfrac{ik_i\exp(-\xi k)}{k^2}(\xi k+(1-2\nu)),\quad i=1,2 \qquad\text{(A.10b)}$$

$$\widetilde{G}_{3j}(k_1,k_2,\xi)=-\dfrac{1+\nu}{2\pi Y}\cdot\dfrac{ik_j\exp(-\xi k)}{k^2}(\xi k-(1-2\nu)),\quad j=1,2 \qquad\text{(A.10c)}$$

$$\tilde{G}_{33}(k_1,k_2,\xi) = \frac{1+\nu}{2\pi Y} \frac{\exp(-\xi k)}{k}(2(1-\nu)+\xi k) \tag{A.10d}$$

where $k \equiv \sqrt{k_1^2 + k_2^2}$. The Fourier representation $\tilde{G}_{ij,l}(k_1,k_2,\xi)$ of the Green's function gradient $\partial G_{ij}/\partial \xi_l$ can be found from Eq. (A.9) as:

$$\tilde{G}_{ij,l}(k_1,k_2,\xi) \equiv \begin{cases} ik_l \tilde{G}_{ij}(k_1,k_2,\xi), & l=1,2 \\ \dfrac{\partial}{\partial \xi}\tilde{G}_{ij}(k_1,k_2,\xi), & l=3 \end{cases} \tag{A.11}$$

Using the Fourier transform of the electric field distribution (Appendix B)

$$E_k(\xi_1,\xi_2,\xi_3) = \frac{1}{2\pi}\int_{-\infty}^{\infty} d\tilde{k}_1 \int_{-\infty}^{\infty} d\tilde{k}_2 \exp(-i\tilde{k}_1 \xi_1 - i\tilde{k}_2 \xi_2)\cdot \tilde{E}_k(\tilde{k}_1,\tilde{k}_2,\xi_3) \tag{A.12}$$

Eq. (A.7) for displacement can be rewritten as:

$$u_i(x_1,x_2) = \int_{-\infty}^{\infty} dk_1 \int_{-\infty}^{\infty} dk_2 \exp(-ik_1 x_1 - ik_2 x_2)\cdot \int_0^{\infty} d\xi\, \tilde{G}_{ij,l}(k_1,k_2,\xi)\tilde{E}_k(k_1,k_2,\xi)e_{klj} \tag{A.13}$$

The displacement is given by a threefold integral in this representation, which however has much simpler structure than the initial Eq. (A.7). Moreover, Eq. (A.13) can be used to calculate the displacement field for materials with arbitrary symmetry of elastic properties provided that approximate or exact Fourier representation for corresponding elastic Green's function is known.

## Appendix B. Fourier Representation for Electric field components.

Here, we derive the representation for the electric field induced by a point charge $Q$ located at a distance $d$ above the surface of the anisotropic half-space with the dielectric permittivity tensor $\varepsilon_{ij}$. The potential distributions below the surface, $V(\mathbf{x})$ (at $x_3 \geq 0$), and above the surface, $V_0(\mathbf{x})$ (at $x_3 < 0$), are obtained from the solution of Laplace's equations:

$$\begin{cases} \varepsilon_0 \varepsilon_{ij} \dfrac{\partial^2}{\partial x_i \partial x_j} V(\mathbf{x}) = 0, \quad x_3 \geq 0 \\ \varepsilon_0 \Delta V_0(\mathbf{x}) = -Q \cdot \delta(x_3 + d)\delta(x_2)\delta(x_1), \quad x_3 < 0 \\ \varepsilon_{3j} \dfrac{\partial V(\mathbf{x})}{\partial x_j}\bigg|_{x_3=0} = \dfrac{\partial V_0(\mathbf{x})}{\partial x_3}\bigg|_{x_3=0}, \quad V(x_3=0) = V_0(x_3=0) \end{cases} \quad (B.1)$$

Here, we introduce Fourier transforms

$$V(\mathbf{x}) = \frac{1}{2\pi} \int_{-\infty}^{\infty} dk_1 \int_{-\infty}^{\infty} dk_2 \exp(-ik_1 x_1 - ik_2 x_2) \cdot \tilde{V}(k_1, k_2, x_3),$$

$$V_0(\mathbf{x}) = \frac{1}{2\pi} \int_{-\infty}^{\infty} dk_1 \int_{-\infty}^{\infty} dk_2 \exp(-ik_1 x_1 - ik_2 x_2) \cdot \tilde{V}_0(k_1, k_2, x_3)$$

and rewrite Eq. (B.1) in Fourier space as:

$$\begin{cases} \left( \varepsilon_{33} \dfrac{\partial^2}{\partial x_3^2} - 2i(\varepsilon_{31} k_1 + \varepsilon_{32} k_2) \dfrac{\partial}{\partial x_3} - \varepsilon_{11} k_1^2 - 2\varepsilon_{12} k_1 k_2 - \varepsilon_{22} k_2^2 \right) \tilde{V} = 0, \quad x_3 \geq 0 \\ \left( \dfrac{\partial^2}{\partial x_3^2} - k_1^2 - k_2^2 \right) \tilde{V}_0 = -\dfrac{Q}{2\pi\varepsilon_0} \cdot \delta(x_3 + d), \quad x_3 < 0 \\ \left( -i\varepsilon_{31} k_1 - i\varepsilon_{32} k_2 + \varepsilon_{33} \dfrac{\partial}{\partial x_3} \right) \tilde{V} = \dfrac{\partial}{\partial x_3} \tilde{V}_0(x_3 = 0), \quad \tilde{V}(x_3 = 0) = \tilde{V}_0(x_3 = 0) \end{cases} \quad (B.2)$$

Where the Fourier transform for Dirac-delta function is $\delta(x) = \dfrac{1}{2\pi} \int_{-\infty}^{\infty} dx \exp(-ikx)$.

The solution of Eq. (B.2) for $x_3 < 0$ can be found in the form $\tilde{V}_0(x_3) = C_0(k_1,k_2)\exp(-k|x_3+d|) + A_0(k_1,k_2)\exp(kx_3)$, where $k \equiv \sqrt{k_1^2 + k_2^2}$. The constant $C_0(k_1,k_2)$ satisfies the equation:

$$(\partial^2/\partial x_3^2 - k^2)\tilde{V}_0(x_3) = -2kC_0(k_1,k_2)\exp(-k\cdot|x_3+d|)\delta(x_3+d) = -\delta(x_3+d)Q/2\pi\varepsilon_0 \quad (B.3)$$

And thus, $C_0 = Q/(4\pi\varepsilon_0 k)$.

The solution of Eq. (B.2) at $x_3 \geq 0$ can be found in the form $\tilde{V}(x_3) = B(k_1,k_2)\exp(-\lambda x_3)$, where the characteristic equation is

$$\varepsilon_{33}\lambda^2 + 2i(\varepsilon_{31}k_1 + \varepsilon_{32}k_2)\lambda - \varepsilon_{11}k_1^2 - 2\varepsilon_{12}k_1k_2 - \varepsilon_{22}k_2^2 = 0. \quad (B.4)$$

The root with positive real part is:

$$\lambda(\mathbf{k}) = \frac{-i(\varepsilon_{31}k_1 + \varepsilon_{32}k_2) + \sqrt{(\varepsilon_{11}\varepsilon_{33} - \varepsilon_{31}^2)k_1^2 - 2(\varepsilon_{31}\varepsilon_{32} - \varepsilon_{12}\varepsilon_{33})k_1k_2 + (\varepsilon_{22}\varepsilon_{33} - \varepsilon_{32}^2)k_2^2}}{\varepsilon_{33}} \quad (B.5)$$

The square root in Eq. (B.5) is real for any real $\mathbf{k} = (k_1, k_2)$ since the matrix of static susceptibility $\varepsilon_{ij}$ is positively defined.

The constants $B(k_1,k_2)$ and $A_0(k_1,k_2)$ satisfy the system of equations:

$$\begin{cases} B(k_1,k_2) = \dfrac{Q}{4\pi\varepsilon_0 k}\exp(-kd) + A_0(k_1,k_2) \\ -(i\varepsilon_{31}k_1 + i\varepsilon_{32}k_2 + \varepsilon_{33}\lambda(\mathbf{k}))B(k_1,k_2) = -\dfrac{Q}{4\pi\varepsilon_0}\exp(-kd) + kA_0(k_1,k_2) \end{cases} \quad (B.6)$$

And the solution of (B.6) yields

$$\begin{aligned} A_0(k_1,k_2) &= \frac{Q\exp(-kd)}{4\pi\varepsilon_0 k}\left(\frac{k - (i\varepsilon_{31}k_1 + i\varepsilon_{32}k_2 + \varepsilon_{33}\lambda(\mathbf{k}))}{k + (i\varepsilon_{31}k_1 + i\varepsilon_{32}k_2 + \varepsilon_{33}\lambda(\mathbf{k}))}\right), \\ B(k_1,k_2) &= \frac{2Q\exp(-kd)}{4\pi\varepsilon_0(k + (i\varepsilon_{31}k_1 + i\varepsilon_{32}k_2 + \varepsilon_{33}\lambda(\mathbf{k})))}. \end{aligned} \quad (B.7)$$

Thus, potential distribution in the Fourier space is given by

$$\tilde{V}(k_1,k_2,x_3) = \frac{2Q\exp(-kd-\lambda(\mathbf{k})x_3)}{4\pi\varepsilon_0(k+(i\varepsilon_{31}k_1+i\varepsilon_{32}k_2+\varepsilon_{33}\lambda(\mathbf{k})))} \tag{B.8}$$

and in real space

$$V(\mathbf{x}) = \frac{1}{2\pi}\int_{-\infty}^{\infty}dk_1\int_{-\infty}^{\infty}dk_2 \frac{2Q\exp(-kd-\lambda(\mathbf{k})x_3)\exp(-ik_1x_1-ik_2x_2)}{4\pi\varepsilon_0\left(k+\sqrt{(\varepsilon_{11}\varepsilon_{33}-\varepsilon_{31}^2)k_1^2-2(\varepsilon_{31}\varepsilon_{32}-\varepsilon_{12}\varepsilon_{33})k_1k_2+(\varepsilon_{22}\varepsilon_{33}-\varepsilon_{32}^2)k_2^2}\right)} \tag{B.9}$$

From Eq. (B.9), the electric field $E_k(\mathbf{x}) = -\partial V(\mathbf{x})/\partial x_k$ in Fourier space $\tilde{E}_k(k_x,k_y,x_3)$ can be found as

$$\tilde{E}_i(k_1,k_2,x_3) \equiv \begin{cases} ik_i\tilde{V}(k_1,k_2,x_3), & i=1,2 \\ -\dfrac{\partial}{\partial x_3}\tilde{V}(k_1,k_2,x_3), & i=3 \end{cases} \tag{B.10}$$

The general expression (B.9) for fully anisotropic dielectric material can be significantly simplified in the case of a transversally isotropic material ($\varepsilon_{ij} = \varepsilon_{ii}\delta_{ij}$, $\varepsilon_{11} = \varepsilon_{22} \neq \varepsilon_{33}$), as discussed in Section III.

## Appendix C. Displacement calculations.

Here, we derive the components of the displacement field for the special case of material with weak dielectric and elastic anisotropy and full piezoelectric anisotropy. Since the piezoelectric tensor $e_{klj}$ is symmetrical on transposition of the indexes $l$ and $j$, Eq. (A.7) can be rewritten as

$$u_i(\mathbf{x}) = W_{ijlk}(\mathbf{x}) e_{klj} \tag{C.1}$$

where components $W_{ijlk}(\mathbf{x})$ are

$$W_{ijlk}(\mathbf{x}) = \int_0^\infty d\xi_3 \int_{-\infty}^\infty d\xi_3 \int_{-\infty}^\infty d\xi_1 \frac{1}{2}\left(\frac{\partial G_{ij}(\mathbf{x},\boldsymbol{\xi})}{\partial \xi_l} + \frac{\partial G_{il}(\mathbf{x},\boldsymbol{\xi})}{\partial \xi_j}\right) E_k(\boldsymbol{\xi}) \tag{C.2a}$$

Or, in Fourier representation,

$$W_{ijlk}(\mathbf{x}) = \int_{-\infty}^\infty dk_1 \int_{-\infty}^\infty dk_2 \exp(-ik_1 x_1 - ik_2 x_2) \cdot \int_0^\infty d\xi \frac{1}{2}\left(\tilde{G}_{ij,l}(k_1,k_2,\xi) + \tilde{G}_{il,j}(k_1,k_2,\xi)\right) \tilde{E}_k(k_1,k_2,\xi) \tag{C.2b}$$

Tensor $W_{ijlk}(\mathbf{x})$ is symmetrical only on the transposition of indexes $l$ and $j$ ($W_{ijlk}(\mathbf{x}) \equiv W_{iljk}(\mathbf{x})$), and thus has 54 nontrivial components in the general case.

When integrating Eq. (C.2), it is convenient to introduce polar coordinates $k_1 \equiv k\cos(\varphi)$, $k_2 \equiv k\sin(\varphi)$. The integration on $\xi$ and $k$ can be done analytically, since the expressions (A.10), (A.11), (B.8), and (B.10) elementary depend on these coordinates. Thus, Eq. (C.2) is reduced to onefold integral on $\varphi$ which can be expressed in terms of elliptic integrals for general dielectric anisotropy. For the case of the transversely isotropic media, these integrals are taken in elementary functions (see Appendix D).

Here we consider the integrate $W_{ijlk}(\mathbf{x})$ analytically for the case of a material with weak dielectric ($\varepsilon_{ij} \approx \kappa \delta_{ij}$) and elastic anisotropy. For clarity, we introduce $x_1 = x$, $x_2 = y$, $\rho^2 = x^2 + y^2$ and $a = \sqrt{x^2 + y^2 + d^2}$.

The elements $W_{ijlk}(\mathbf{x})$ containing indexes "1" and/or "2", can be obtained one from another by simultaneous permutation of indexes "1" ↔ "2" and coordinates x ↔ y, e.g.,

$W_{1111}(x,y) \equiv W_{2222}(y,x)$,  $W_{1223}(x,y) \equiv W_{2113}(y,x)$,  $W_{2211}(x,y) \equiv W_{1122}(y,x)$,

$W_{2311}(x,y) \equiv W_{1322}(y,x)$, $W_{2333}(x,y) \equiv W_{1333}(y,x)$. Therefore, the number of nontrivial components of $W_{ijlk}(\mathbf{x})$ is reduced to 28. The components determining the displacement component $u_1$ are the following:

$$W_{1111}(x,y) = -\frac{Q}{2\pi\varepsilon_0(\kappa+1)} \frac{1+\nu}{2\pi Y} \frac{\pi}{4a(a+d)^2 \rho^4} \times$$
$$\times \begin{pmatrix} d(5a+2d)x^4 + 2(4a^2 - ad + 4d^2)x^2 y^2 + a(2a+5d)y^4 + \\ + 2(1-2\nu)(d(a+2d)x^4 + 6ad\,x^2 y^2 + a(2a+d)y^4) \end{pmatrix}$$
(C.3a)

$$W_{1112}(x,y) = \frac{Q}{2\pi\varepsilon_0(\kappa+1)} \frac{1+\nu}{2\pi Y} \frac{\pi x y}{2a(a+d)^3 \rho^2} \times$$
$$\times \left((4a+d)x^2 + (a+4d)y^2 + 2(1-2\nu)(d\,x^2 + a\,y^2)\right)$$
(C.3b)

$$W_{1113}(x,y) = \frac{Q}{2\pi\varepsilon_0(\kappa+1)} \frac{1+\nu}{2\pi Y} \frac{\pi x}{4a(a+d)^2 \rho^2} \times$$
$$\times \left((5a+2d)x^2 + (-a+8d)y^2 + 2(1-2\nu)((a+2d)x^2 + 3a\,y^2)\right)$$
(C.3c)

$$W_{1121}(x,y) = \frac{Q}{2\pi\varepsilon_0(\kappa+1)} \frac{1+\nu}{2\pi Y} \frac{\pi x y}{2a(a+d)^3 \rho^2} \times$$
$$\times \left((2a-d)x^2 + (-a+2d)y^2 + 2(1-2\nu)(d\,x^2 + a\,y^2)\right)$$
(C.3d)

$$W_{1122}(x,y) = -\frac{Q}{2\pi\varepsilon_0(\kappa+1)}\frac{1+\nu}{2\pi Y}\frac{\pi}{4a(a+d)^2\rho^4} \times$$
$$\times \begin{pmatrix} a(4a+d)x^4 - 2(a^2 - 7ad + d^2)x^2y^2 + d(a+4d)y^4 + \\ + 2(1-2\nu)(adx^4 + 2(a^2 - ad + d^2)x^2y^2 + ady^4) \end{pmatrix} \quad \text{(C.3e)}$$

$$W_{1123}(x,y) = \frac{Q}{2\pi\varepsilon_0(\kappa+1)}\frac{1+\nu}{2\pi Y}\frac{\pi y}{4a(a+d)^2\rho^2} \times$$
$$\times \left((7a-2d)x^2 + (a+4d)y^2 + 2(1-2\nu)((-a+2d)x^2 + ay^2)\right) \quad \text{(C.3f)}$$

$$W_{1221}(x,y) = \frac{Q}{2\pi\varepsilon_0(\kappa+1)}\frac{1+\nu}{2\pi Y}\frac{\pi(adx^4 + 2(a^2 - ad + d^2)x^2y^2 + ady^4)}{4a(a+d)^2\rho^4}(1+4\nu) \quad \text{(C.3g)}$$

$$W_{1222}(x,y) = -\frac{Q}{2\pi\varepsilon_0(\kappa+1)}\frac{1+\nu}{2\pi Y}\frac{\pi xy(ax^2 + dy^2)}{2a(a+d)^3\rho^2}(1+4\nu) \quad \text{(C.3h)}$$

$$W_{1223}(x,y) = -\frac{Q}{2\pi\varepsilon_0(\kappa+1)}\frac{1+\nu}{2\pi Y}\frac{\pi x(ax^2 + (-a+2d)y^2)}{4a(a+d)^2\rho^2}(1+4\nu) \quad \text{(C.3i)}$$

$$W_{1131}(x,y) = -\frac{Q}{2\pi\varepsilon_0(\kappa+1)}\frac{1+\nu}{2\pi Y}\frac{\pi x((3a+2d)x^2 + (a+4d)y^2)}{4a(a+d)^2\rho^2} \quad \text{(C.3j)}$$

$$W_{1132}(x,y) = -\frac{Q}{2\pi\varepsilon_0(\kappa+1)}\frac{1+\nu}{2\pi Y}\frac{\pi y((5a+2d)x^2 + (3a+4d)y^2)}{4a(a+d)^2\rho^2} \quad \text{(C.3k)}$$

$$W_{1133}(x,y) = -\frac{Q}{2\pi\varepsilon_0(\kappa+1)}\frac{1+\nu}{2\pi Y}\frac{\pi((2a+d)x^2 + (a+2d)y^2)}{2a(a+d)\rho^2} \quad \text{(C.3l)}$$

$$W_{1231}(x,y) = \frac{Q}{2\pi\varepsilon_0(\kappa+1)}\frac{1+\nu}{2\pi Y}\frac{\pi y((-a+2d)x^2 + ay^2)}{4a(a+d)^2\rho^2} \quad \text{(C.3m)}$$

$$W_{1232}(x,y) = \frac{Q}{2\pi\varepsilon_0(\kappa+1)}\frac{1+\nu}{2\pi Y}\frac{\pi x(ax^2 + (-a+2d)y^2)}{4a(a+d)^2\rho^2} \quad \text{(C.3n)}$$

$$W_{1233}(x,y) = -\frac{Q}{2\pi\varepsilon_0(\kappa+1)}\frac{1+\nu}{2\pi Y}\frac{\pi xy}{2a(a+d)^2} \quad \text{(C.3o)}$$

$$W_{1331}(x,y) = \frac{Q}{2\pi\varepsilon_0(\kappa+1)}\frac{1+\nu}{2\pi Y}\frac{\pi(dx^2 + ay^2)}{2a(a+d)\rho^2}(-1+4\nu) \quad \text{(C.3p)}$$

$$W_{1332}(x,y) = -\frac{Q}{2\pi\varepsilon_0(\kappa+1)}\frac{1+\nu}{2\pi Y}\frac{\pi x y}{2a(a+d)^2}(-1+4\nu) \qquad (C.3q)$$

$$W_{1333}(x,y) = -\frac{Q}{2\pi\varepsilon_0(\kappa+1)}\frac{1+\nu}{2\pi Y}\frac{\pi x}{2a(a+d)}(-1+4\nu) \qquad (C.3r)$$

Components related to $u_2$ can be obtained from (C.3) with the help of the simultaneous permutation of indexes "1" ↔ "2" and coordinates x ↔ y. Components determining the displacement component $u_3$ are the following:

$$W_{3111}(x,y) = \frac{Q}{2\pi\varepsilon_0(\kappa+1)}\frac{1+\nu}{2\pi Y}\frac{\pi x\left((a+2d)x^2+3ay^2\right)}{4a(a+d)^2\rho^2}(-1+4\nu) \qquad (C.4a)$$

$$W_{3112}(x,y) = \frac{Q}{2\pi\varepsilon_0(\kappa+1)}\frac{1+\nu}{2\pi Y}\frac{\pi y\left((-a+2d)x^2+ay^2\right)}{4a(a+d)^2\rho^2}(-1+4\nu) \qquad (C.4b)$$

$$W_{3113}(x,y) = \frac{Q}{2\pi\varepsilon_0(\kappa+1)}\frac{1+\nu}{2\pi Y}\frac{\pi\left(dx^2+ay^2\right)}{2a(a+d)\rho^2}(-1+4\nu) \qquad (C.4c)$$

$$W_{3121}(x,y) \equiv W_{3112}(x,y) \qquad (C.4b)$$

$$W_{3123}(x,y) = -\frac{Q}{2\pi\varepsilon_0(\kappa+1)}\frac{1+\nu}{2\pi Y}\frac{\pi x y}{2a(a+d)^2}(-1+4\nu) \qquad (C.4d)$$

$$W_{3131}(x,y) = -\frac{Q}{2\pi\varepsilon_0(\kappa+1)}\frac{1+\nu}{2\pi Y}\frac{\pi\left(dx^2+ay^2\right)}{2a(a+d)\rho^2} \qquad (C.4e)$$

$$W_{3132}(x,y) = \frac{Q}{2\pi\varepsilon_0(\kappa+1)}\frac{1+\nu}{2\pi Y}\frac{\pi x y}{2a(a+d)^2} \qquad (C.4f)$$

$$W_{3133}(x,y) = \frac{Q}{2\pi\varepsilon_0(\kappa+1)}\frac{1+\nu}{2\pi Y}\frac{\pi x}{2a(a+d)} \qquad (C.4g)$$

$$W_{3331}(x,y) = -\frac{Q}{2\pi\varepsilon_0(\kappa+1)}\frac{1+\nu}{2\pi Y}\frac{\pi x}{2a(a+d)}(3-4\nu) \qquad (C.4h)$$

$$W_{3333}(x,y) = -\frac{Q}{2\pi\varepsilon_0(\kappa+1)}\frac{1+\nu}{2\pi Y}\frac{\pi}{2a}(3-4\nu) \qquad (C.4i)$$

Here we listed only noncomponents that cannot be found with the help of the above-mentioned rule.

While the integration in Eqs. (C2,3) is performed for elastically isotropic material, these results can be easily generalized for a transversely isotropic material.

# Appendix D. Integration on $\varphi$.

The integration on polar angle $\varphi$ reduces to the following integrals

$$\int_0^{2\pi} \frac{d\varphi}{\alpha + \beta\cos(\varphi) + \chi\sin(\varphi)} = \frac{2\pi}{\sqrt{\alpha^2 - \beta^2 - \chi^2}} \tag{D.1}$$

$$\int_0^{2\pi} \frac{\cos(\varphi)d\varphi}{\alpha + \beta\cos(\varphi) + \chi\sin(\varphi)} = -\frac{2\pi\beta}{\sqrt{\alpha^2 - \beta^2 - \chi^2}} \frac{1}{\sqrt{\alpha^2 - \beta^2 - \chi^2} + \alpha} \tag{D.2}$$

$$\int_0^{2\pi} \frac{\sin(\varphi)d\varphi}{\alpha + \beta\cos(\varphi) + \chi\sin(\varphi)} = -\frac{2\pi\chi}{\sqrt{\alpha^2 - \beta^2 - \chi^2}} \frac{1}{\sqrt{\alpha^2 - \beta^2 - \chi^2} + \alpha} \tag{D.3}$$

$$\int_0^{2\pi} \frac{\cos(\varphi)^2 d\varphi}{\alpha + \beta\cos(\varphi) + \chi\sin(\varphi)} = \frac{2\pi}{(\beta^2 + \chi^2)^2}\left(-\alpha(\beta^2 - \chi^2) + \frac{\alpha^2(\beta^2 - \chi^2) + \chi^2(\beta^2 + \chi^2)}{\sqrt{\alpha^2 - \beta^2 - \chi^2}}\right) \tag{D.4}$$

$$\int_0^{2\pi} \frac{\sin(\varphi)^2 d\varphi}{\alpha + \beta\cos(\varphi) + \chi\sin(\varphi)} = \frac{2\pi}{(\beta^2 + \chi^2)^2}\left(\alpha(\beta^2 - \chi^2) + \frac{-\alpha^2(\beta^2 - \chi^2) + \beta^2(\beta^2 + \chi^2)}{\sqrt{\alpha^2 - \beta^2 - \chi^2}}\right) \tag{D.5}$$

$$\int_0^{2\pi} \frac{\cos(\varphi)\sin(\varphi)d\varphi}{\alpha + \beta\cos(\varphi) + \chi\sin(\varphi)} = \frac{2\pi\beta\chi}{2\alpha(\alpha^2 - \beta^2 - \chi^2) + (2\alpha^2 - \beta^2 - \chi^2)\sqrt{\alpha^2 - \beta^2 - \chi^2}} \tag{D.6}$$

$$\int_0^{2\pi} \frac{\sin(\varphi)^3 d\varphi}{\alpha + \beta\cos(\varphi) + \chi\sin(\varphi)} =$$
$$= \frac{2\pi\chi}{(\beta^2 + \chi^2)^3}\left(-2\alpha^2(3\beta^2 - \chi^2) + (3\beta^2 + \chi^2)(\beta^2 + \chi^2) + \frac{\alpha^3(6\beta^2 - 2\chi^2) - 6\alpha\beta^2(\beta^2 + \chi^2)}{\sqrt{\alpha^2 - \beta^2 - \chi^2}}\right) \tag{D.7}$$

$$\int_0^{2\pi} \frac{\cos(\varphi)^3 d\varphi}{\alpha + \beta\cos(\varphi) + \chi\sin(\varphi)} =$$
$$= \frac{2\pi\beta}{(\beta^2 + \chi^2)^3}\left(2\alpha^2(\beta^2 - 3\chi^2) + (\beta^2 + 3\chi^2)(\beta^2 + \chi^2) + \frac{-\alpha^3(2\beta^2 - 6\chi^2) - 6\alpha\chi^2(\beta^2 + \chi^2)}{\sqrt{\alpha^2 - \beta^2 - \chi^2}}\right) \tag{D.8}$$

It should be noted that (D.1)-(D.8) are obtained for $\alpha > 0$, $\alpha^2 > \beta^2 + \chi^2$. This is the case since $\alpha \equiv d$, $\beta \equiv ix$, $\chi \equiv iy$.

## Appendix E. Tip-surface potential.

Here, we derive the approximations for tip-induced surface potential as a function of tip-surface separation, avoiding the summation over large number ($N > 1000$) of image charges. The tip is represented by a biased conductive sphere of radius $R_0$. Its apex is located at a distance $\Delta R$ from the sample surface. The potential at $z > 0$ is

$$V_Q(\rho) = \frac{Q_0}{2\pi\varepsilon_0(\kappa+1)} \sum_{m=0}^{\infty} \frac{q_m}{\sqrt{\rho^2 + (d-d_m)^2}}, \qquad \text{(E.1a)}$$

Where the magnitude of image charges are

$$q_m = \left(\frac{\kappa-1}{\kappa+1}\right) \cdot \frac{R_0}{2d - r_{m-1}} q_{m-1} \equiv \left(\frac{\kappa-1}{\kappa+1}\right)^m \frac{sh(\theta)}{sh((m+1)\theta)}, \qquad \text{(E.1b)}$$

The corresponding coordinates from the center of the sphere are

$$d_m = \frac{R_0^2}{2d - d_{m-1}} \equiv R_0 \frac{sh(m\theta)}{sh((m+1)\theta)}. \qquad \text{(E.1c)}$$

And $d_0 = 0$, $q_0 = 1$, $ch(\theta) = d/R_0$.

The limiting cases for Eq. (E.1) are evaluated as following:

(a) For large tip-surface separations $R_0 \ll d$, Eq. (E.1) becomes

$$V_Q(\rho) \rightarrow \frac{Q_0}{2\pi\varepsilon_0(\kappa+1)} \frac{1}{\sqrt{\rho^2 + d^2}} \text{ as expected.}$$

(b) On the symmetry axis of the system, $\rho \rightarrow 0$, $d \approx R_0$ and $\theta(m+1) \ll 1$, the image charges $\frac{q_m}{d - d_m} \approx \left(\frac{\kappa-1}{\kappa+1}\right)^m \frac{\exp(-m\theta)}{d}$, so the sum over image charges is

$$\sum_{m=0}^{\infty} \left(\frac{\kappa-1}{\kappa+1}\right)^m \exp(-m\theta) = \left(1 - \left(\frac{\kappa-1}{\kappa+1}\right)\exp(-\theta)\right)^{-1} \text{ and hence}$$

$V_Q(0) \to \dfrac{Q_0}{2\pi\varepsilon_0 d} \dfrac{1}{\kappa(1-\exp(-\theta))+(1+\exp(-\theta))}$. For large separations, the expected asymptotic behavior $V_Q(\rho \gg d) \approx \dfrac{Q_0}{2\pi\varepsilon_0(\kappa-1)}\ln\left(\dfrac{\kappa+1}{2}\right)\dfrac{1}{\sqrt{\rho^2+d^2}}$ at $d = R_0$ and $\rho \gg d$ is recovered

(c) Finally, for $k \to 0$ and $\theta(m+1) \ll 1$ the image charges $q_m \approx \left(\dfrac{\kappa-1}{\kappa+1}\right)^m \dfrac{\exp(-m\theta)}{m+1}$.

Summation $\sum\limits_{m=0}^{\infty}\left(\dfrac{\kappa-1}{\kappa+1}\right)^m \dfrac{\exp(-m\theta)}{m+1} = \dfrac{\kappa+1}{\kappa-1}\exp(\theta)\ln\left(\dfrac{\kappa+1}{\kappa+1-(\kappa-1)\exp(-\theta)}\right)$ yields

$V_Q(\rho \gg d) \to \dfrac{Q_0}{2\pi\varepsilon_0(\kappa-1)} \dfrac{\exp(\theta)}{\sqrt{\rho^2+d^2}}\ln\left(\dfrac{\kappa+1}{\kappa+1-(\kappa-1)\exp(-\theta)}\right)$, in agreement with expected limit $V_Q(\rho \gg d) \approx \dfrac{Q_0}{2\pi\varepsilon_0(\kappa-1)}\ln\left(\dfrac{\kappa+1}{2}\right)\dfrac{1}{\sqrt{\rho^2+d^2}}$ at $d = R_0$ and $\rho \gg d$.

Keeping in mind (a)–(c), we select the more general approximation for a potential on the surface, $z = 0$, induced by sphere, from the exact series (C.1), in the case when there is a small gap between the tip apex and the sample surface. This approximation is

$$V_Q(\rho) \approx \dfrac{Q_0}{2\pi\varepsilon_0}\dfrac{1}{\sqrt{\rho^2+d^2}}\left(\dfrac{1}{\kappa(1-\exp(-\mu\varphi(\rho)))+(1+\exp(-\mu\varphi(\rho)))} + \dfrac{\exp(\theta)}{\kappa-1}\ln\left(\dfrac{\kappa+1}{\kappa+1-(\kappa-1)\exp(-\theta)}\right)\dfrac{\rho}{\sqrt{\rho^2+d^2}}\right),$$ (E.2)

$ch\varphi(\rho) = \dfrac{\sqrt{\rho^2+d^2}}{R_0}$, $ch(\theta) = \dfrac{d}{R_0}$, $0 < \mu \leq 1$ is a fitting parameter

From Eq. (E.2), the potential below the tip is $V_Q(0) \approx \dfrac{Q_0}{2\pi\varepsilon_0}\dfrac{1}{d}f\left(\dfrac{d}{R_0}\right)$, where

$f(x) = \dfrac{1}{\kappa(1-\exp(-\mu\,arcchx))+(1+\exp(-\mu\,arcchx))}$ and $f(1) = 0.5$. Taking into account that

the charge $Q_0 = 4\pi\varepsilon_0 R_0 U$, one obtains that $V_Q(0) = U$ at $R_0 = d$. Shown in Fig. E-1. is a comparison of exact and approximate solutions.

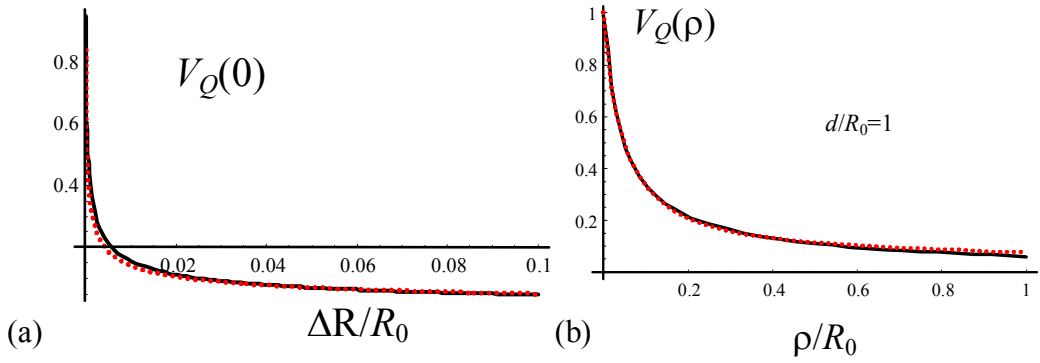

**Fig. E-1.** Comparison of exact solution Eq. (E1) (solid curve) and approximation by Eq. (E.2) for μ=0.5 (dotted curve). Shown are (a) potential below the tip as a function of tip-surface separation and (b) surface potential as a function of distance from the point of contact.

[28] N.N. Lebedev, I.P. Skal'skaya, and Ya.S. Uflyand, *Problems in Mathematical Physics*, Pergamon Press, 1966.

[29] Y.V. Sharvin, Zh. Eksp. Teor. Fiz. **48**, 984 (1965).

[30] R. Proksch, private communications.

[31] P. De Wolf, Veeco, presentation at the UK-SPM 2006 meeting.

[32] A.G. Onaran, M. Balantekin, W. Lee, W.L. Hughes, B.A. Buchine, R.O. Guldiken, Z. Parlak, C.F. Quate, and F.L. Degertekin, Rev. Sci. Instrum. **77**, 023501 (2006).

[33] M. Dienwiebel, E. de Kuyper, L. Crama, J. W. M. Frenken, J. A. Heimberg, D.-J. Spaanderman, D. Glastra van Loon, T. Zijlstra, and E. van der Drift, Rev. Sci. Instrum. **76**, 043704 (2005).

[34] L.M. Eng, H.-J. Guntherodt, G.A. Schneider, U. Kopke, and J.M. Saldana, Appl. Phys. Lett. **74**, 233 (1999).

[35] B.J. Rodriguez, A.Gruverman, A.I. Kingon, R.J. Nemanich, and J.S. Cross, J. Appl. Phys. **95**, 1958 (2004).

[36] S.V. Kalinin, E. Karapetian, and M. Kachanov, Phys. Rev. B **70**, 184101 (2004).

[37] E. Karapetian, M. Kachanov, and S.V. Kalinin, Phil. Mag. **85**, 1017 (2005).

[38] U. D. Schwarz, J. Colloid Interface Sci. **261**, 99 (2003).

[39] H.J. Butt, B. Cappella, and M. Kappl, Surf. Sci. Reports **59**, 1 (2005).

[40] C.S. Ganpule, V. Nagarjan, H. Li, A.S. Ogale, D.E. Steinhauer, S. Aggarwal, E. Williams, R. Ramesh, and P. De Wolf, Appl. Phys. Lett. **77**, 292 (2000).

[41] A. Agronin, M. Molotskii, Y. Rosenwaks, E. Strassburg, A. Boag, S. Mutchnik, and G. Rosenman, J. Appl. Phys. **97**, 084312 (2005).

[42] F. Felten, G.A. Schneider, J. Muñoz Saldaña, and S.V. Kalinin, J. Appl. Phys. **96**, 563 (2004).

[43] D.A. Scrymgeour and V. Gopalan, Phys. Rev. **B 72**, 024103 (2005).

[44] S.V. Kalinin, E.A. Eliseev, and A.N. Morozovska, Appl. Phys. Lett. **88**, 232904 (2006).

[45] J.D. Jackson, *Classical Electrodynamics*, John Wiley, New York, 1998.

[46] E.J. Mele, Am. J. Phys. **69**, 557 (2001).

[47] T. Mura, *Micromechanics of Defects in Solids*, Martinus Nijhoff, Dordrecht, 1987.

[48] M. Kachanov, B. Shafiro, and I. Tsukrov, *Handbook of Elasticity Solutions*, Springer 2003.

[49] A. I. Lurie, *Spatial problems of the elasticity theory*, (Gos. Izd. Teor. Tekh. Lit., Moscow, 1955).

[50] L.D. Landau and E.M. Lifshitz, *Theory of Elasticity*. Theoretical Physics, Vol. 7 (Butterworth-Heinemann, Oxford, U.K., 1998).

[51] K.L. Johnson, Contact Mechanics, Cambridge Univ. Press, Cambridge, UK, 1985.

[52] S. Belaidi, P. Girard, and G. Leveque, J. Appl. Phys. **81**, 1023 (1997).

[53] Note that factors "2", "4" are introduced in transition to Voigt notations for elastic compliances tensor $\hat{s}$ and piezoelectric stress tensor $\hat{d}$, while for elastic stiffnesses $\hat{c}$ and piezoelectric tensor $\hat{e}$ there are no factors in the notations due to the relationship $e_{kij} = d_{klm} c_{lmij}$ [see e.g., J.F. Nye, *Physical Properties of Crystals*. New York: Oxford University Press (1985)].

[54] M. Abplanalp, Dr. Nat. Sci. thesis, Swiss Federal Institute of Technology, Zurich, 2001.

[55] S.V. Kalinin, B.J. Rodriguez, J. Shin, S. Jesse, V. Grichko, T. Thundat, A.P. Baddorf, and A. Gruverman, Ultramicroscopy **106**, 334 (2006).